\documentclass[12pt,prd,tightenlines,nofootinbib]{revtex4}
\usepackage{bm}
\usepackage{graphics}
\usepackage{rotating}
\usepackage{epsfig}
\usepackage{slashed}
\begin{document}
\title{
Rare $\Lambda_b\to\Lambda l^+l^-$ and $\Lambda_b\to\Lambda\gamma$ decays in
the relativistic quark model}  
\author{R. N. Faustov}
\author{V. O. Galkin}
\affiliation{Institute of Informatics in Education, FRC CSC RAS,
  Vavilov Street 40, 119333 Moscow, Russia}

\begin{abstract}
Rare $\Lambda_b\to\Lambda l^+l^-$ and $\Lambda_b\to\Lambda\gamma$
decays are investigated in the relativistic quark model based on the
quark-diquark picture of baryons. The decay form factors are
calculated with the account of all relativistic effects including
relativistic transformations of baryon wave functions from rest to
moving reference frame and the contribution of the intermediate negative energy states. The
momentum transfer  squared dependence of the form factors is
explicitly determined in the whole accessible kinematical range. The
calculated decay branching fractions, various forward-backward
asymmetries for the rare decay $\Lambda_b\to\Lambda \mu^+\mu^-$  are
found to be consistent with recent detailed measurements by the LHCb
Collaboration. Predictions for the $\Lambda_b\to\Lambda \tau^+\tau^-$
decay observables are given.  
\end{abstract}


\maketitle

\section{Introduction}
\label{sec:int}

In the standard model the exclusive rare weak decays of hadrons,
governed by the $b\to
s$ quark transitions,  proceed through the flavour changing neutral
currents. Therefore they provide a sensitive test of different
new physics extensions (see e.g. Ref.~\cite{aab} and references
therein). Such transitions were studied in detail both theoretically
and experimentally in the $B$ and $B_s$ meson decays. Differential
decay rates and angular distributions, various asymmetry parameters
were measured with rather high accuracy \cite{bgh}.  However, several
tensions between experimental data and the standard model predictions
have been found \cite{aab,bgh,lhcb1}.  It was argued \cite{lhcb1}  that
these differences could be explained by contributions from physics
beyond the standard model, or by unexpectedly large hadronic
effects that are not properly accounted for in the predictions.
Therefore it is important to refine theoretical description of the
rare $b\to s$ transitions and search for similar effects in other rare
decays. 

The rare semileptonic $\Lambda_b \to\Lambda l^+l^-$ and
radiative $\Lambda_b\to\Lambda\gamma$ decays provide a valuable testing
ground. Indeed, the first observation of the baryonic flavour changing
neutral current decay $\Lambda_b\to\Lambda\mu^+\mu^-$ was reported in 2011 by
the CDF Collaboration \cite{cdf}. It was followed by more
comprehensive and precise data from the LHCb Collaboration
\cite{lhcb2013,lhcb2015}, which not only measured the differential
branching fraction but also performed the detailed angular analysis of the
$\Lambda_b\to\Lambda\mu^+\mu^-$ decay. Theoretical description of such decays
is based on the effective Hamiltonian in which intermediate gauge
bosons are integrated out. The short and long distance contributions are
separated by application of the operator product expansion. The short
distance effects are described by the Wilson coefficients, while the
calculation of the long distance part involves  consideration of the hadronic
matrix elements of the corresponding weak currents between baryon
states, which are usually parametrized by the set of invariant form factors.
The kinematically accessible momentum transfer squared $q^2$ range in
such decays is rather broad. However, most of the theoretical approaches available in
the literature provide determination of the decay form factors only in
one particular kinematical point or in the limited range. Thus
light-cone QCD sum rules determine form factors at large recoil of
the final hadron (near $q^2=0$) while the lattice QCD calculations are reliable in the
small recoil region (near $q^2=q^2_{\rm max}$). Then extrapolation of the
theoretical predictions to the whole kinematical range is
needed which introduces additional theoretical uncertainties. Therefore, reliable determination of the
$q^2$ dependence of the hadronic form factors in the whole kinematical
range without extrapolations or  model assumptions is
important for increasing the reliability of theoretical predictions.

In this paper we apply the relativistic quark model \cite{mmass} based on the
quasipotential approach with the QCD-motivated interquark interaction to the calculation of the matrix elements of
the flavour changing neutral current between baryon states. In our
model  baryons are considered to be the relativistic quark-diquark
bound systems. Their wave functions are known from the baryon mass
spectra calculations \cite{bmass}. 
Let us note that at present the convincing evidence  of the existence of diquark
correlations in hadrons has been collected. Information comes from
different sectors of hadron physics. Thus, in the light meson sector
it has been argued for a long time \cite{jaffe} that mesons forming the inverted
lightest scalar  nonet can be well described as 
tetraquarks   treated as diquark-antidiquark bound states \cite{ltetr}.
In the heavy meson sector several charged charmonium- and
bottomonium-like states were discovered \cite{hq}. They should be inevitably
multiquark, at least four quark --- tetraquark, states. One of the
most successful pictures of such tetraquark states is the
diquark-antidiquark model \cite{hq,htetr}.  In the baryon
sector it is well known that the number of observed excited states
both in the
light and heavy sectors is considerably lower than the number of
excited states predicted in the three-quark approach. The introduction
of diquarks significantly reduces this number, since in such a picture some of degrees of freedom are frozen
and thus the number of possible excitations is substantially
smaller. The calculations of the heavy and strange baryon spectra
\cite{bmass} show that all available experimental data can be well
described in the framework of the relativistic quark-diquark picture
of baryons. The lattice QCD calculations indicate existence of the
diquark correlations in baryons \cite{lattb}. Very recently the Belle Collaboration \cite{belle} published
data on the production cross sections of charmed baryons in $e^+e^-$
annihilation. The observed a factor of three excess of the
production cross section of $\Lambda_c$ states over $\Sigma_c$ states provides  a strong support for a diquark structure in
the ground state and low-lying excited $\Lambda_c$ baryons.
  
 Calculating the weak
current matrix elements between baryon states we systematically take
into account all relativistic corrections including contributions of
the intermediate negative energy states and relativistic
transformations of baryon wave functions from rest to the moving
reference frame using the methods previously developed for the
description of the semileptonic baryon decays \cite{sllbdecay}. Such an
approach allows us to obtain explicitly the $q^2$ dependence of the form factors in the whole kinematical
range. For our calculation we use the same effective Hamiltonian and
Wilson coefficients as in our previous consideration of the rare $B$
and $B_s$ meson decays \cite{rareB}.  

The paper is organized as follows. In Sec.~\ref{sec:mod} we briefly
describe our relativistic quark-diquark model of baryons, present the
relevant quasipotential equation and give expressions for the weak
current matrix element. The rare $b\to s$ transition form factors are
calculated in Sec.~\ref{sec:ff}. They are expressed through the
overlap integrals of the baryon wave functions. The analytic
expressions for the form factors, which
accurately reproduce numerical results in the whole accessible kinematical
$q^2$ range,  are presented. Comparison with the previous
calculations for the form factor values at $q^2=0$ and $q^2=q^2_{\rm
  max}$ is given. Rare semileptonic $\Lambda_b\to \Lambda l^+l^-$
decays are considered in Sec.~\ref{sec:rd}. Differential branching
fractions and other angular observables are calculated and compared
with available experimental data and lattice calculations. The
estimates for the rare radiative $\Lambda_b\to\Lambda\gamma$ decay are
presented in Sec.~\ref{sec:rrad}. Our conclusions are given in
Sec.~\ref{sec:concl}, while the  Appendix contains explicit expressions for the
rare decay form factors.

\section{Relativistic quark-diquark model}
\label{sec:mod}

For the calculation of the rare $\Lambda_b$ baryon decays we use the same
relativistic quark-diquark model which was previously employed for the
calculation of the baryon masses \cite{bmass} and weak semileptonic
decays \cite{sllbdecay}. The initial $\Lambda_b$ and final
$\Lambda$ baryons are considered as the bound states of the heavy $Q$ ($b$ or
$s$) quark and light scalar $[u,d]$ diquark. They are described by the
wave function $\Psi_{\Lambda_Q}$, which satisfy the relativistic
quasipotential equation of the Schr\"odinger type \cite{bmass}
\begin{equation}
\label{quasipot}
{\left(\frac{b^2(M)}{2\mu_{R}}-\frac{{\bf
p}^2}{2\mu_{R}}\right)\Psi_{\Lambda_Q}({\bf p})} =\int\frac{d^3 q}{(2\pi)^3}
 V({\bf p,q};M)\Psi_{\Lambda_Q}({\bf q}),
\end{equation}  
where the relativistic reduced mass and  and the center-of-mass system
relative momentum squared on the mass shell are given by
$$
\mu_{R}=\frac{M_{\Lambda_Q}^4-(m^2_Q-m^2_d)^2}{4M_{\Lambda_Q}^3}, $$
$${b^2(M) }
=\frac{[M_{\Lambda_Q}^2-(m_Q+m_d)^2][M_{\Lambda_Q}^2-(m_Q-m_d)^2]}{4M_{\Lambda_Q}^2}.
$$
Here $M_{\Lambda_Q}$, $m_Q$ and $m_d$ are the $\Lambda_Q$ baryon mass,
$Q$ quark mass and  diquark $d$ mass, respectively. The quark-diquark
interaction quasipotential $V({\bf p,q};M)$ (see explicit expressions
in Ref.~\cite{bmass}) is the relativistic
generalization of the Cornell potential
\begin{equation}
V(r)=-\frac43\frac{\tilde\alpha_s(r)}{r}+Ar+B,
\end{equation}
where the first term is the smeared Coulomb potential with
$\tilde\alpha_s(r)\equiv\alpha_s F(r)$ and the form factor $F(r)$ takes the
diquark internal structure into account \cite{bmass}.  The confining quark interaction was taken to be the mixture of the
Lorentz-vector and scalar linearly growing with $r$  potentials
\begin{equation}
V^V_{\rm conf}(r)=(1-\varepsilon)(Ar+B),\qquad\
V^S_{\rm conf}(r) =\varepsilon (Ar+B),
\end{equation}
 with the mixing coefficient $\varepsilon$, which was set to
 $\varepsilon=-1$ from the consideration of meson properties  \cite{mmass}.
The vertex of the long-range vector quark interaction contains not
only the Dirac part but an additional Pauli term, thus introducing the
long-range anomalous chromomagnetic quark moment $\kappa$. Its value
was set to $\kappa=-1$ in our previous consideration of meson
properties \cite{mmass}. Such choice provides the vanishing of the long-range chromomagnetic
contribution to the potential, which is proportional to $(1+\kappa)$.       
 The constituent quark masses $m_{u,d}=0.33$
GeV, $m_s=0.5$ GeV, $m_c=1.55$ GeV, $m_b=4.88$ GeV and the parameters of the linear potential
$A=0.18$ GeV$^2$ and $B=-0.3$ GeV have the usual values of quark
models. The mass of the scalar $[u,d]$ diquark was calculated to be
$m_d=0.71$~GeV \cite{bmass}.

\begin{figure}
  \centering
  \includegraphics{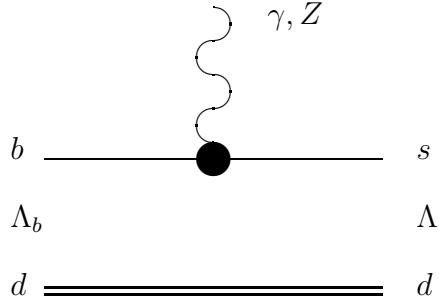}
\caption{Lowest order vertex function $\Gamma^{(1)}$
contributing to the current matrix element. \label{d1}}
\end{figure}

\begin{figure}
  \centering
  \includegraphics{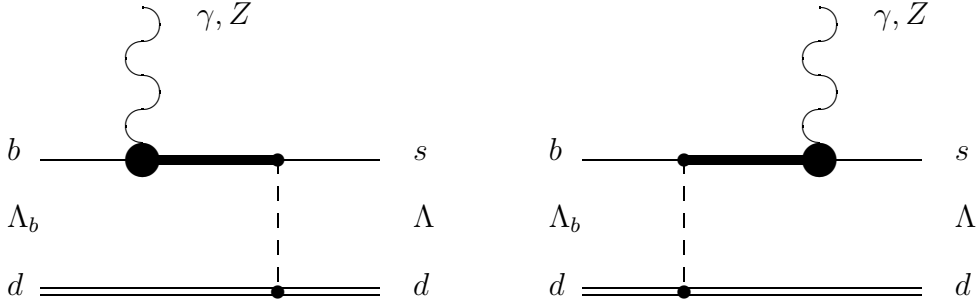}
\caption{ Vertex function $\Gamma^{(2)}$
taking the quark interaction into account. Dashed lines correspond  
to the effective quark-diquark quasipotential ${\cal V}$. Bold lines
denote the negative-energy part of the quark propagator. \label{d2}}
\end{figure}

The matrix element of the weak current $J^W$, governing the rare $b\to s
l^+l^-$ transition,  between baryon
states  in the considered approach is given by \cite{f,sllbdecay}
\begin{equation}\label{mel} 
\langle \Lambda(P) \vert J^W_\mu \vert \Lambda_b(Q)\rangle
=\int \frac{d^3p\, d^3q}{(2\pi )^6} \bar \Psi_{\Lambda\,{\bf P}}({\bf
p})\Gamma _\mu ({\bf p},{\bf q})\Psi_{\Lambda_b\,{\bf Q}}({\bf q}),
\end{equation}
where $\Gamma _\mu ({\bf p},{\bf
q})$ is the two-particle vertex function. It receives relativistic
contributions both from the impulse approximation diagram in Fig.~\ref{d1}
\begin{equation}
\Gamma_\mu^{(1)}({\bf
p},{\bf q})=\psi^*_d(p_d)\bar
u_{s}(p_{s})\gamma_\mu(1-\gamma^5)u_b(q_b)\psi_d(q_d) 
(2\pi)^3\delta({\bf p}_d-{\bf
q}_d),\end{equation}
 and from the
diagrams with the intermediate negative-energy states in Fig.~\ref{d2} which are the consequence
of the projection onto the positive-energy states in the
quasipotential approach
\begin{eqnarray}\label{gamma2} 
\Gamma_\mu^{(2)}({\bf
p},{\bf q})&=& \psi^*_d(p_d)\bar u_{s}(p_{s}) \Bigl\{\gamma_{\mu}(1-\gamma^5)
\frac{\Lambda_b^{(-)}(
k)}{\epsilon_b(k)+\epsilon_b(p_{s})}\gamma^0
{\cal V}({\bf p}_d-{\bf
q}_d)\nonumber \\ 
& &+{\cal V}({\bf p}_d-{\bf
q}_d)\frac{\Lambda_{s}^{(-)}(k')}{ \epsilon_{s}(k')+
\epsilon_{s}(q_b)}\gamma^0 \gamma_{\mu}(1-\gamma^5)\Bigr\}u_b(q_b)
\psi_d(q_d).\end{eqnarray}  
Here $\psi_d(p)$ is the diquark wave
function; ${\cal V}({\bf p})$ is the quark-diquark  interaction quasipotential;
 ${\bf k}={\bf p}_{s}-{\bf\Delta};\
{\bf k}'={\bf q}_b+{\bf\Delta};\ {\bf\Delta}=
{\bf P}-{\bf Q}$; \ $\epsilon(p)=\sqrt{m^2+{\bf p}^2}$; and
$$\Lambda^{(-)}(p)=\frac{\epsilon(p)-\bigl( m\gamma
^0+\gamma^0({\bm{ \gamma}{\bf p}})\bigr)}{ 2\epsilon (p)}.$$
$\Psi_{\Lambda\,{\bf P}}({\bf p})$ is the baryon  wave function projected onto the positive-energy states of 
quarks and boosted to the moving reference frame with momentum ${\bf
  P}$ \cite{f,sllbdecay}
\begin{equation}
\label{wig}
\Psi_{{\Lambda}\,{\bf P }}({\bf
p})=D_{q}^{1/2}(R_{L_{\bf P}}^W)D_d(R_{L_{
\bf P}}^W)\Psi_{{\Lambda}\,{\bf 0}}({\bf p}),
\end{equation}
where $\Psi_{{\Lambda}\,{\bf 0}}$ is the baryon wave function in the
rest frame,  $R^W$ is the Wigner rotation, $L_{\bf P}$ is the Lorentz boost
from the baryon rest frame to a moving one with momentum ${\bf P}$, and   
 $D^{1/2}_q(R^W)$ is the rotation matrix of the quark spin \cite{sllbdecay}, while 
 the rotation matrix for the scalar diquark spin $D_d(R^W)=1$.

\section{Form factors  of the rare $\Lambda_b$ baryon decays}

\label{sec:ff}

The matrix element of the flavour changing neutral current for the
rare $\Lambda_b\to\Lambda l^+l^-$ baryon decay can be
parametrized by the following set of invariant form factors  
\begin{eqnarray}
  \label{eq:ff}
  \langle \Lambda(p',s')|\bar{s} \gamma^\mu b|\Lambda_b(p,s)\rangle&=& \bar
  u_{\Lambda}(p',s')\Bigl[f_1^V(q^2)\gamma^\mu-f_2^V(q^2)i\sigma^{\mu\nu}\frac{q_\nu}{M_{\Lambda_b}}+f_3^V(q^2)\frac{q^\mu}{M_{\Lambda_b}}\Bigl]
u_{\Lambda_b}(p,s),\cr
\!\!\!\! \langle \Lambda(p',s')|\bar{s} \gamma^\mu\gamma_5 b|\Lambda_b(p,s)\rangle&=& \bar
  u_{\Lambda}(p',s')[f_1^A(q^2)\gamma^\mu-f_2^A(q^2)i\sigma^{\mu\nu}\frac{q_\nu}{M_{\Lambda_b}}+f_3^A(q^2)\frac{q^\mu}{M_{\Lambda_b}}\Bigl]
\gamma_5 u_{\Lambda_b}(p,s),\qquad \cr
\langle \Lambda(p',s') | \bar{s} i\sigma^{\mu\nu}q_\nu b | \Lambda_b(p,s) \rangle &=& \bar{u}_\Lambda(p',s') \left[  \frac{f_1^{TV}(q^2)}{M_{\Lambda_b}} \left(\gamma^\mu q^2 - q^\mu \slashed{q} \right) - f_2^{TV}(q^2) i\sigma^{\mu\nu}q_\nu  \right] u_{\Lambda_b}(p,s), \cr
\!\!\!\!\!\!\! \langle \Lambda(p',s') | \bar{s} i\sigma^{\mu\nu}q_\nu\gamma_5 b | \Lambda_b(p,s) \rangle &=& \bar{u}_\Lambda(p',s') \left[  \frac{f_1^{TA}(q^2)}{M_{\Lambda_b}} \left(\gamma^\mu q^2 - q^\mu \slashed{q} \right) - f_2^{TA}(q^2) i\sigma^{\mu\nu}q_\nu  \right]\gamma_5 u_{\Lambda_b}(p,s), 
\end{eqnarray}
where   $u_{\Lambda_{b}}(p,s)$ and
$u_{\Lambda}(p',s')$ are Dirac spinors of the initial and final
baryon; $q=p'-p$.

The other popular parameterization is the helicity-based definition of the form factors from Ref.~\cite{fy}

\begin{eqnarray}
\label{eq:hff}
 \langle \Lambda(p',s') | \bar{s} \gamma^\mu b | \Lambda_b(p,s) \rangle &=&
 \bar{u}_\Lambda(p',s') \bigg[ f_0(q^2) (M_{\Lambda_b}-M_\Lambda)\frac{q^\mu}{q^2} \cr
 && + f_+(q^2) \frac{M_{\Lambda_b}+M_\Lambda}{s_+}\left( p^\mu + p^{\prime \mu} - (M_{\Lambda_b}^2-M_\Lambda^2)\frac{q^\mu}{q^2}  \right) \cr
 && + f_\perp(q^2) \left(\gamma^\mu - \frac{2M_\Lambda}{s_+} p^\mu -
    \frac{2 M_{\Lambda_b}}{s_+} p^{\prime \mu} \right) \bigg]
    u_{\Lambda_b}(p,s), \cr
 \langle \Lambda(p',s') | \bar{s} \gamma^\mu\gamma_5 b | \Lambda_b(p,s) \rangle &=&
 -\bar{u}_\Lambda(p',s') \:\gamma_5 \bigg[ g_0(q^2)\: (M_{\Lambda_b}+M_\Lambda)\frac{q^\mu}{q^2} \cr
  && + g_+(q^2)\frac{M_{\Lambda_b}-M_\Lambda}{s_-}\left( p^\mu + p^{\prime \mu} - (M_{\Lambda_b}^2-M_\Lambda^2)\frac{q^\mu}{q^2}  \right) \cr
 && + g_\perp(q^2) \left(\gamma^\mu + \frac{2M_\Lambda}{s_-} p^\mu - \frac{2 M_{\Lambda_b}}{s_-} p^{\prime \mu} \right) \bigg]  u_{\Lambda_b}(p,s), \cr
 \langle \Lambda(p',s') | \bar{s} i\sigma^{\mu\nu} q_\nu  b | \Lambda_b(p,s) \rangle &=&
 - \bar{u}_\Lambda(p',s') \bigg[  h_+(q^2) \frac{q^2}{s_+} \left( p^\mu + p^{\prime \mu} - (M_{\Lambda_b}^2-M_{\Lambda}^2)\frac{q^\mu}{q^2} \right) \cr
 &&  + h_\perp(q^2)\, (M_{\Lambda_b}+M_\Lambda) \left( \gamma^\mu -
    \frac{2  M_\Lambda}{s_+} \, p^\mu - \frac{2M_{\Lambda_b}}{s_+} 
    p^{\prime \mu}   \right) \bigg] u_{\Lambda_b}(p,s), \cr
\!\!\!\! \langle \Lambda(p',s')| \bar{s}  i\sigma^{\mu\nu}q_\nu \gamma_5   b|\Lambda_b(p,s)\rangle &=&
 -\bar{u}_{\Lambda}(p',s') \, \gamma_5 \bigg[   \tilde{h}_+(q^2) \, \frac{q^2}{s_-} \left( p^\mu + p^{\prime \mu} -  (M_{\Lambda_b}^2-M_{\Lambda}^2) \frac{q^\mu}{q^2} \right) \cr
 &&   + \tilde{h}_\perp(q^2)\,  (M_{\Lambda_b}-M_\Lambda) \left(
    \gamma^\mu +  \frac{2 M_\Lambda}{s_-} \, p^\mu - \frac{2
    M_{\Lambda_b}}{s_-} \, p^{\prime \mu}  \right) \bigg]
    u_{\Lambda_b}(p,s),\ \ \ \ 
\end{eqnarray}
with $s_\pm =(M_{\Lambda_b} \pm M_\Lambda)^2-q^2$.

The form factors (\ref{eq:hff}) and (\ref{eq:ff}) are related in the
following way
\begin{eqnarray}
 f_+(q^2)&=& f_1^V(q^2) + \frac{q^2}{M_{\Lambda_b}(M_{\Lambda_b}+M_\Lambda)} f_2^V(q^2),  \cr
 f_\perp(q^2)&=& f_1^V(q^2) + \frac{M_{\Lambda_b}+M_\Lambda}{M_{\Lambda_b}} f_2^V(q^2),  \cr
 f_0(q^2)&=& f_1^V(q^2) + \frac{q^2}{M_{\Lambda_b}(M_{\Lambda_b}-M_\Lambda)} f_3^V(q^2), \cr
 g_+(q^2)&=& f_1^A(q^2) - \frac{q^2}{M_{\Lambda_b}(M_{\Lambda_b}-M_\Lambda)} f_2^A(q^2), \cr
 g_\perp(q^2)&=& f_1^A(q^2) - \frac{M_{\Lambda_b}-M_\Lambda}{M_{\Lambda_b}} f_2^A(q^2), \cr
 g_0(q^2)&=& f_1^A(q^2) - \frac{q^2}{M_{\Lambda_b}(M_{\Lambda_b}+M_\Lambda)} f_3^A(q^2), \cr
 h_+(q^2)&=& -f_2^{TV}(q^2) - \frac{M_{\Lambda_b}+M_\Lambda}{M_{\Lambda_b}} f_1^{TV}(q^2),  \cr
 h_\perp(q^2)&=& -f_2^{TV}(q^2) - \frac{q^2}{M_{\Lambda_b}(M_{\Lambda_b}+M_\Lambda)} f_1^{TV}(q^2),  \cr
 \tilde{h}_+(q^2)&=& -f_2^{TA}(q^2) + \frac{M_{\Lambda_b}-M_\Lambda}{M_{\Lambda_b}} f_1^{TA}(q^2),  \cr
 \tilde{h}_\perp(q^2)&=& -f_2^{TA}(q^2) + \frac{q^2}{M_{\Lambda_b}(M_{\Lambda_b}-M_\Lambda)} f_1^{TA}(q^2). 
\end{eqnarray}

To find the weak decay form factors we need to calculate the  matrix element of the weak current between baryon wave functions known from the mass spectra calculations.
The expressions for the decay form factors $f_i^{V,A}$ ($i=1,2,3$)  parameterizing matrix elements
of the vector and axial vector weak currents between baryon states
were obtained in our previous paper \cite{sllbdecay}. They are given
in the Appendix of Ref.~\cite{sllbdecay}. For the calculation of the rare
baryon decays we need to extend our analysis and get expressions for
the rest of form factors  $f_i^{TV,TA}$ parameterizing matrix elements
of the tensor and pseudo tensor currents. To achieve this goal we
follow the approach developed in Ref.~\cite{sllbdecay}.  Namely we use the
$\delta$-function in the expression for the lowest-order vertex function
$\Gamma^{(1)}$ arising in the impulse approximation (see Fig. 1)  to express
the current matrix element (\ref{mel}) as the usual overlap integral of baryon
wave functions. Thus, this contribution can be calculated exactly  in the
whole kinematical range. On the other hand, the consideration of the vertex function
$\Gamma^{(2)}$ (see Fig. 2) is more
complicated, since this function takes into account
contributions coming from the negative-energy parts of the quark
propagators and thus explicitly depends on the quark-diquark  potential, in particular, on the Lorentz-structure of
the confining interaction.   Taking into account that the recoil
momentum of the final $\Lambda$ baryon $|{\bf\Delta}|$, in the initial
$\Lambda_b$ baryon rest frame, is
significantly larger than the relative quark momentum in the baryon
almost in the whole accessible kinematical range,\footnote{The square
  of the momentum transfer squared to the lepton pair $q^2$ varies  from 0 to
  $q^2_{\rm  max}\approx 20$~GeV$^2$ for the $\Lambda_b$ decays to $\Lambda$.} we neglect small
relative momentum $|{\bf p}|$ with respect to the recoil momentum $|{\bf\Delta}|$ in the energies of
quarks composing the energetic final $\Lambda$ baryon and replace $\epsilon_{q}(p+\Delta)\equiv\sqrt{m_{q}^2+({\bf 
p}+{\bf\Delta})^2} $ with $\epsilon_{q}(\Delta)\equiv
\sqrt{m_{q}^2+{\bf\Delta}^2}$. As a result we can use the
quasipotential equation to take one of the integrations in the current
matrix element (\ref{mel}) and again get the expression for the current matrix
element as the usual overlap integral of baryon wave functions. It is
important to point out that such an approach allows us to
consistently take into account all relativistic corrections including boosts of the baryon wave functions from the
rest frame to the moving one (\ref {wig}) and contributions of the intermediate
negative-energy states.   The obtained
expressions for the form factors are presented in the Appendix (to
simplify these expressions, as previously, we explicitly set the long-range anomalous
chromomagnetic quark moment $\kappa=-1$).

\begin{figure}
\centering
  \includegraphics[width=8cm]{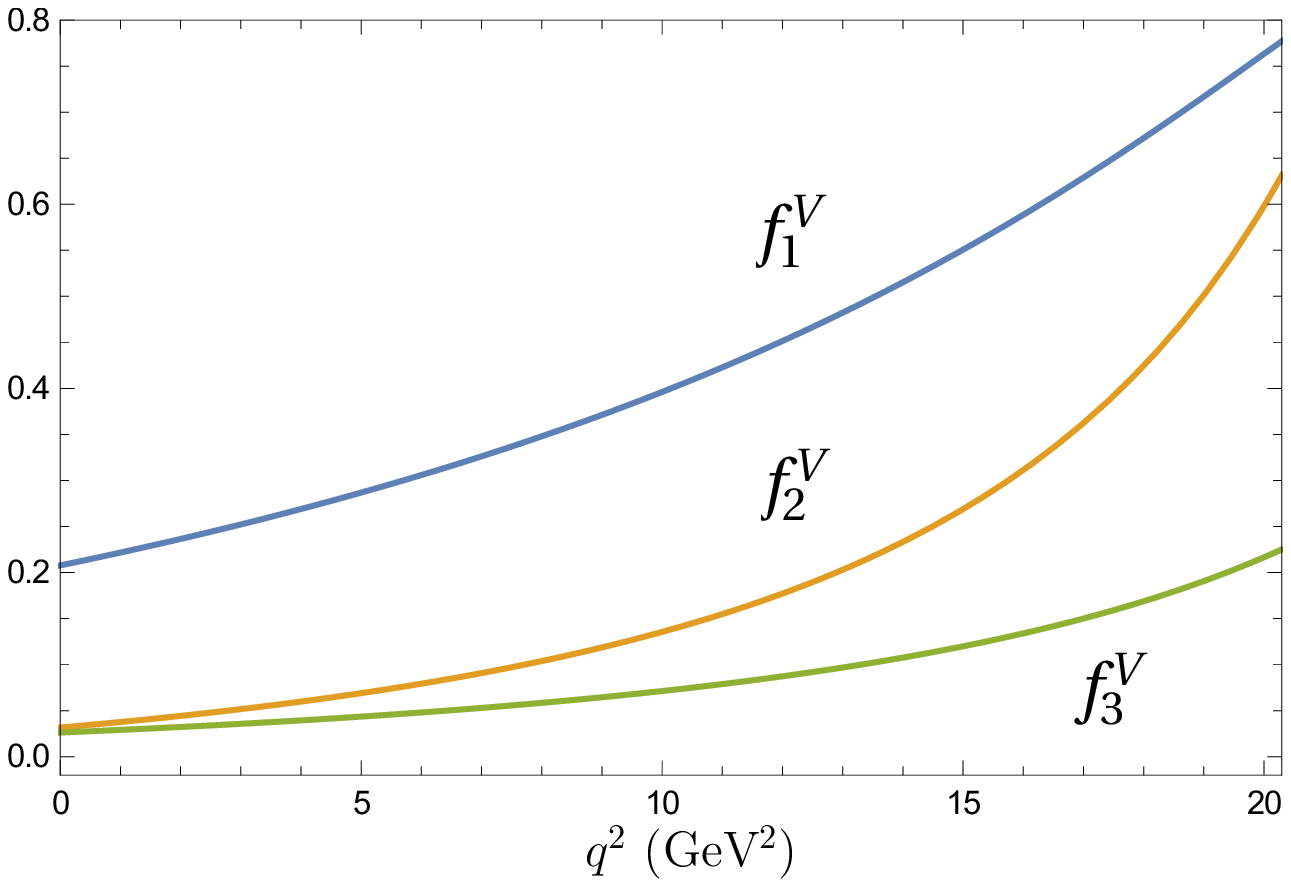}\ \
 \ \includegraphics[width=8cm]{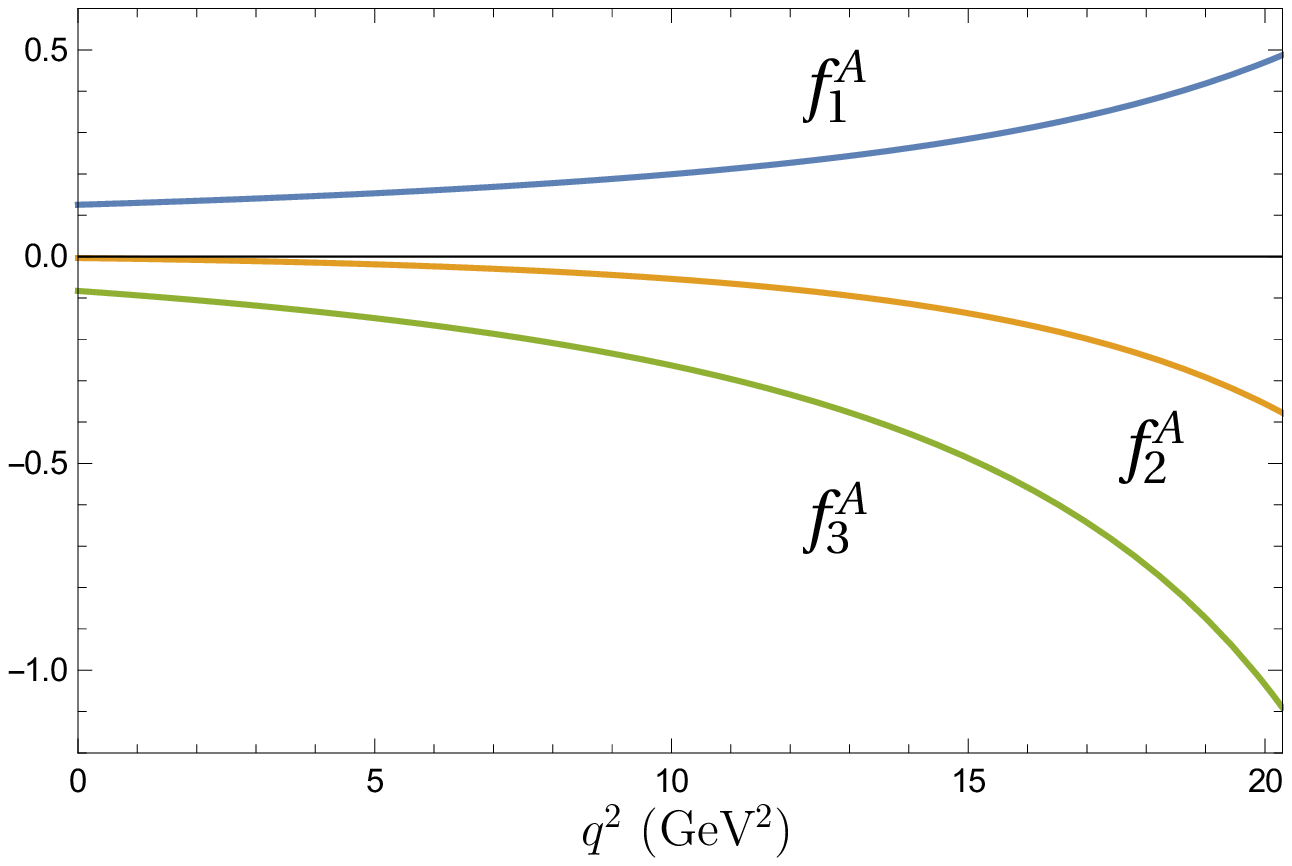}\\
\includegraphics[width=8cm]{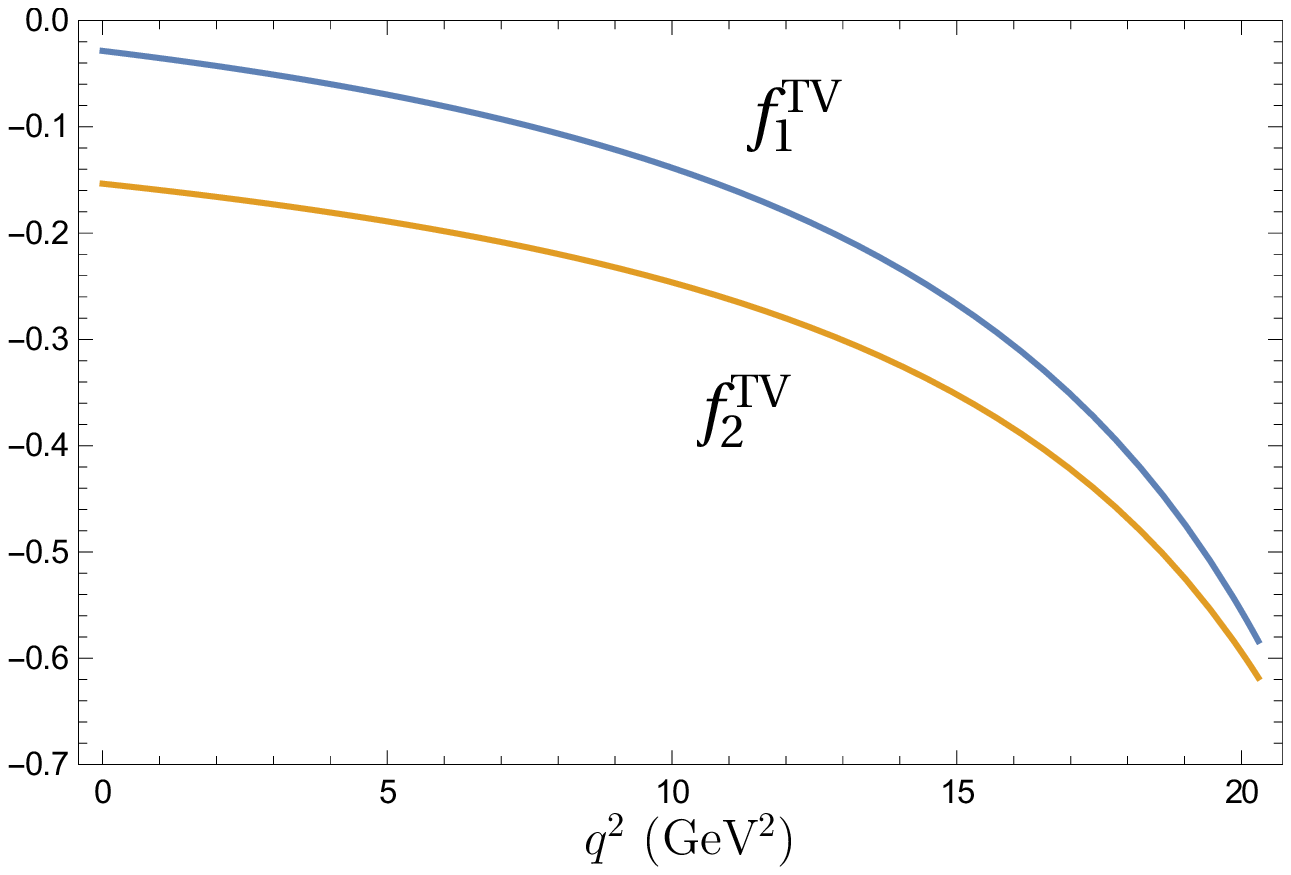}\ \
 \ \includegraphics[width=8cm]{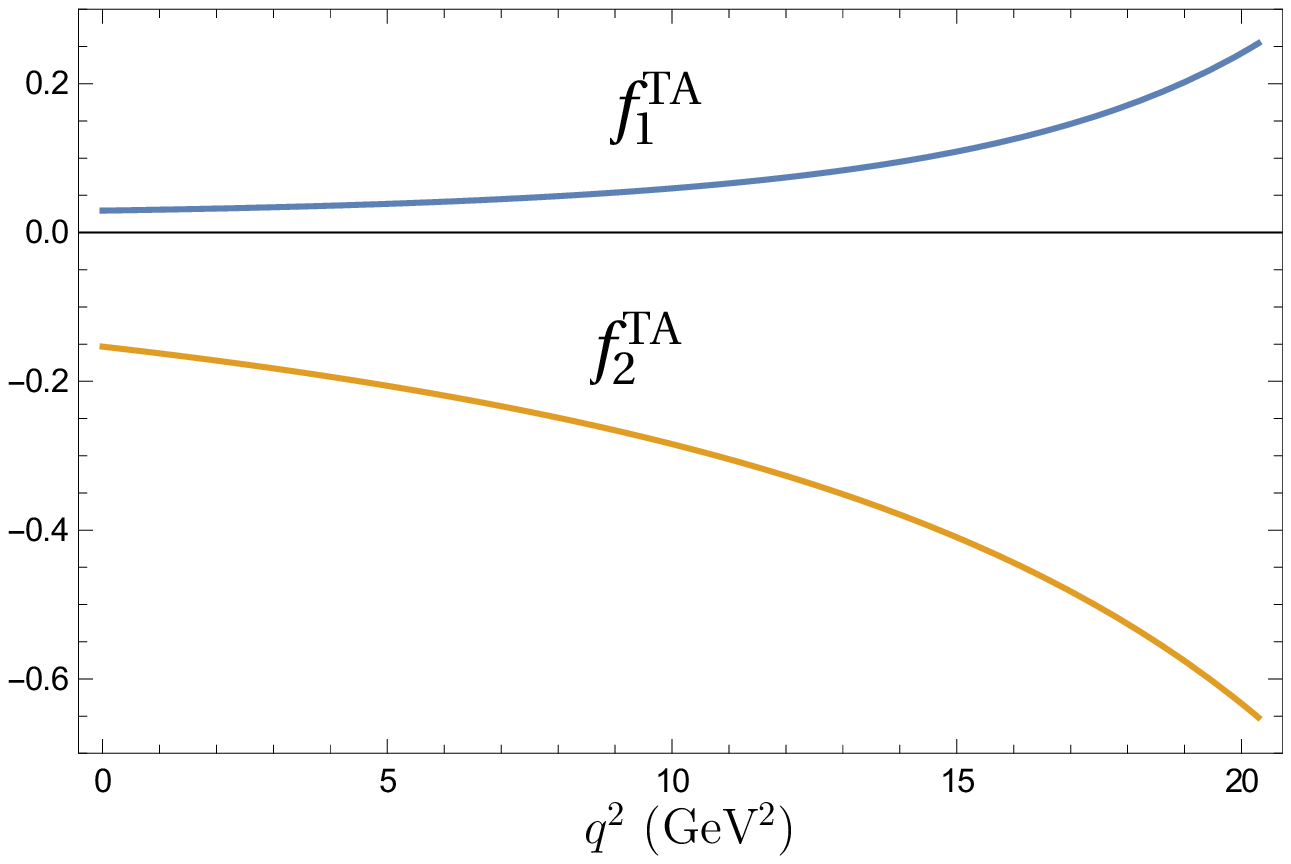}\\
\caption{Form factors of the rare weak $\Lambda_b\to \Lambda$ transition.    } 
\label{fig:ffLbLs}
\end{figure}

Substituting the baryon wave functions, found into the calculation of
their mass spectra, in the expressions for the decay form factors we
calculate their values and explicitly determine their 
dependence on the momentum transfer  squared $q^2$  in the whole kinematical range.  We find that the weak decay form factors can 
be approximated with a high accuracy by the  expressions: 
\begin{equation}
  \label{fitff}
  F(q^2)= \frac{1}{\displaystyle{1-\frac{q^2}{M_{\rm pole}^2}}} \left\{ a_0 + a_1 z(q^2) +
    a_2 [z(q^2)]^2 \right\},
\end{equation}
where the variable 
\begin{equation}
z(q^2) = \frac{\sqrt{t_+-q^2}-\sqrt{t_+-t_0}}{\sqrt{t_+-q^2}+\sqrt{t_+-t_0}}.
\end{equation}
Following Ref.~\cite{latt} we take  $t_+=(M_B+M_K)^2$ and $t_0 = q^2_{\rm max} = (M_{\Lambda_b} - M_{\Lambda})^2$.  The pole
masses have the  values: $M_{\rm
  pole}\equiv M_{B_s^*}=5.416$ GeV for $f_{1,2}^V$, $f_{1,2}^{TV}$; $M_{\rm
  pole}\equiv M_{B_{s1}}=5.830$ GeV for $f_{1,2}^A$, $f_{1,2}^{TA}$; $M_{\rm
  pole}\equiv M_{B_{s0}}=5.833$ GeV for $f_{3}^V$;  $M_{\rm
  pole}\equiv M_{B_{s}}=5.366$ GeV for $f_{3}^A$.
The fitted values of the parameters $a_0$, $a_1$, $a_2$ as well as the
values of form factors at maximum $q^2=0$ and zero recoil $q^2=q^2_{\rm
  max}$ are given in Table~\ref{ffLbL}. The difference of the fitted
form factors from the calculated ones does not exceed 0.5\%. Our model form factors
are plotted in Fig.~\ref{fig:ffLbLs}.

\begin{table}
\caption{Calculated form factors of the rare weak $\Lambda_b\to \Lambda$ transition. }
\label{ffLbL}
\begin{ruledtabular}
\begin{tabular}{ccccccccccc}
& $f^V_1(q^2)$ & $f^V_2(q^2)$& $f^V_3(q^2)$& $f^A_1(q^2)$ & $f^A_2(q^2)$ &$f^A_3(q^2)$& $f^{TV}_1(q^2)$ & $f^{TV}_2(q^2)$& $f^{TA}_1(q^2)$ & $f^{TA}_2(q^2)$\\
\hline
$f(0)$          &0.208 &$0.032$ & $0.026$ & 0.125 & $-0.003$&$-0.083$&$-0.029$ & $-0.153$ & 0.029 & $-0.153$\\
$f(q^2_{\rm max})$&0.777  &$0.632$ & $0.225$ & 0.487& $-0.378$& $-1.09$ &$-0.584$ & $-0.618$ & 0.254& $-0.652$\\
$a_0$      &$0.239$&$0.195$& $0.091$& $0.196$ &$-0.152$&  $-0.322$&$-0.180$& $-0.190$& $0.102$ &$-0.263$\\
$a_1$      &$0.633$&$-0.765$&$-0.280$&$-0.515$&$1.03$&$1.05$&$0.540$&$0.252$&$-0.592$&$0.351$\\
$a_2$      &$-3.34$&$0.252$& $0$& $0.899$ &$-1.68$&  $-0.033$&$0.505$& $-0.398$& $1.19$ &$0.531$\\
\end{tabular}
\end{ruledtabular}
\end{table}

\begin{figure}
\centering
  \includegraphics[width=8cm]{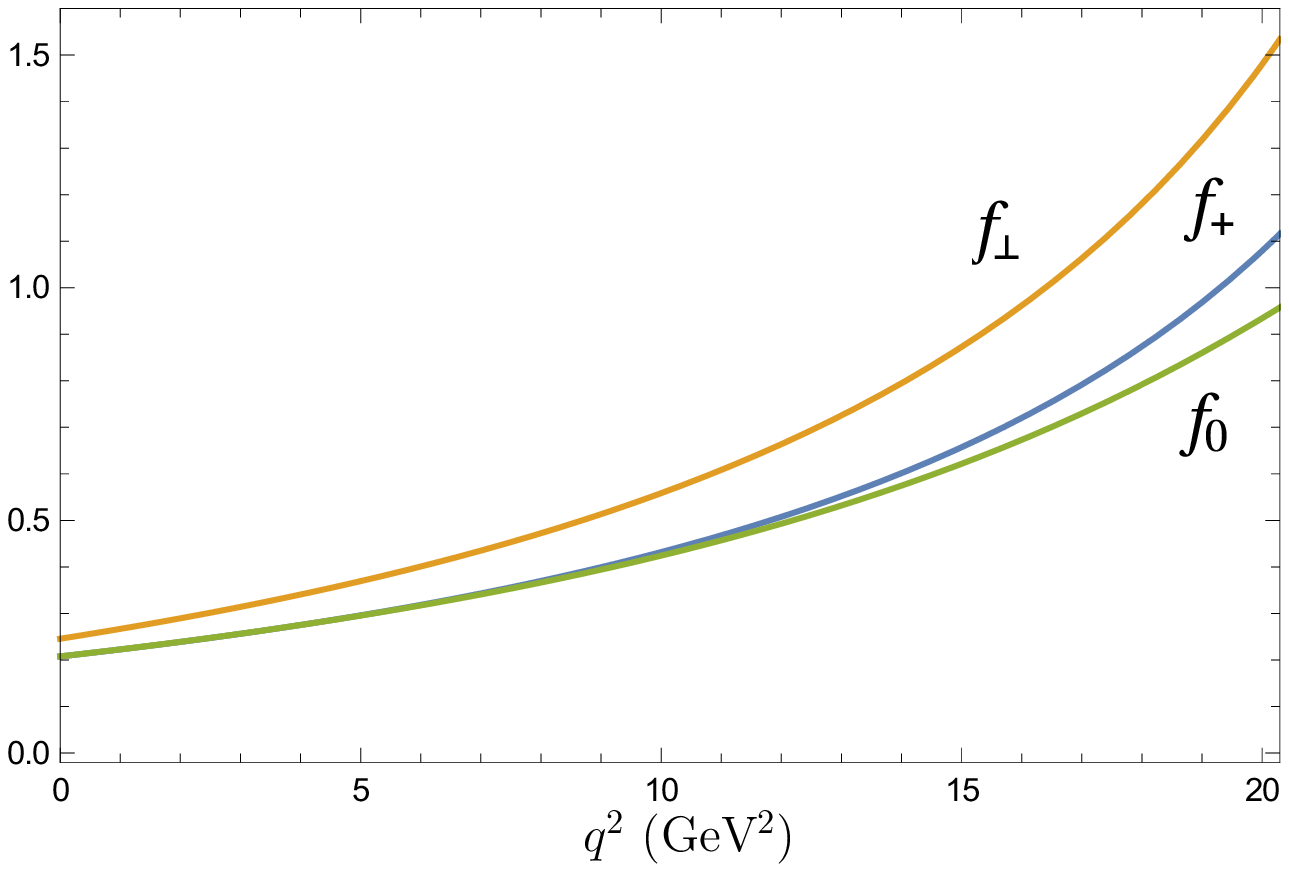}\ \
 \ \includegraphics[width=8cm]{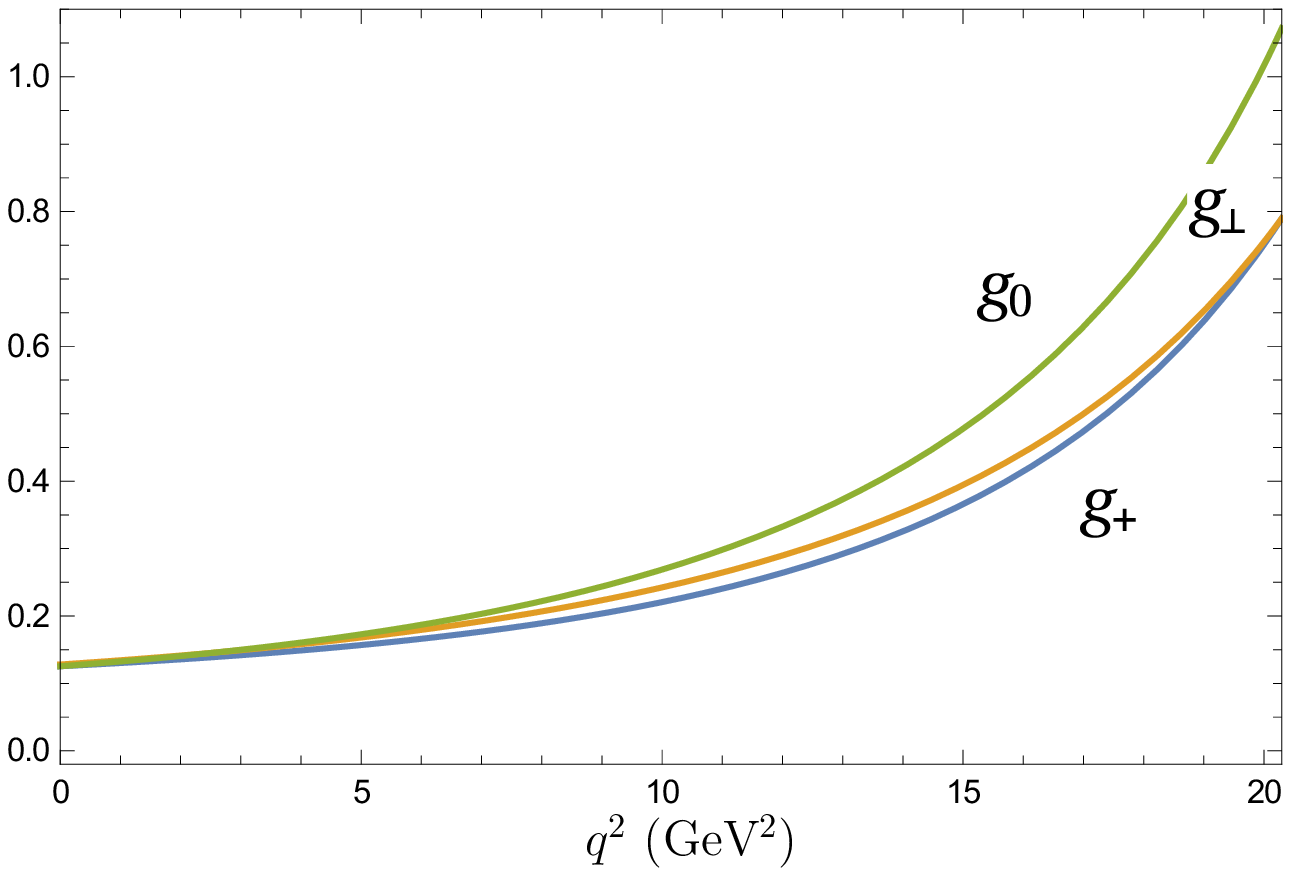}\\
\includegraphics[width=8cm]{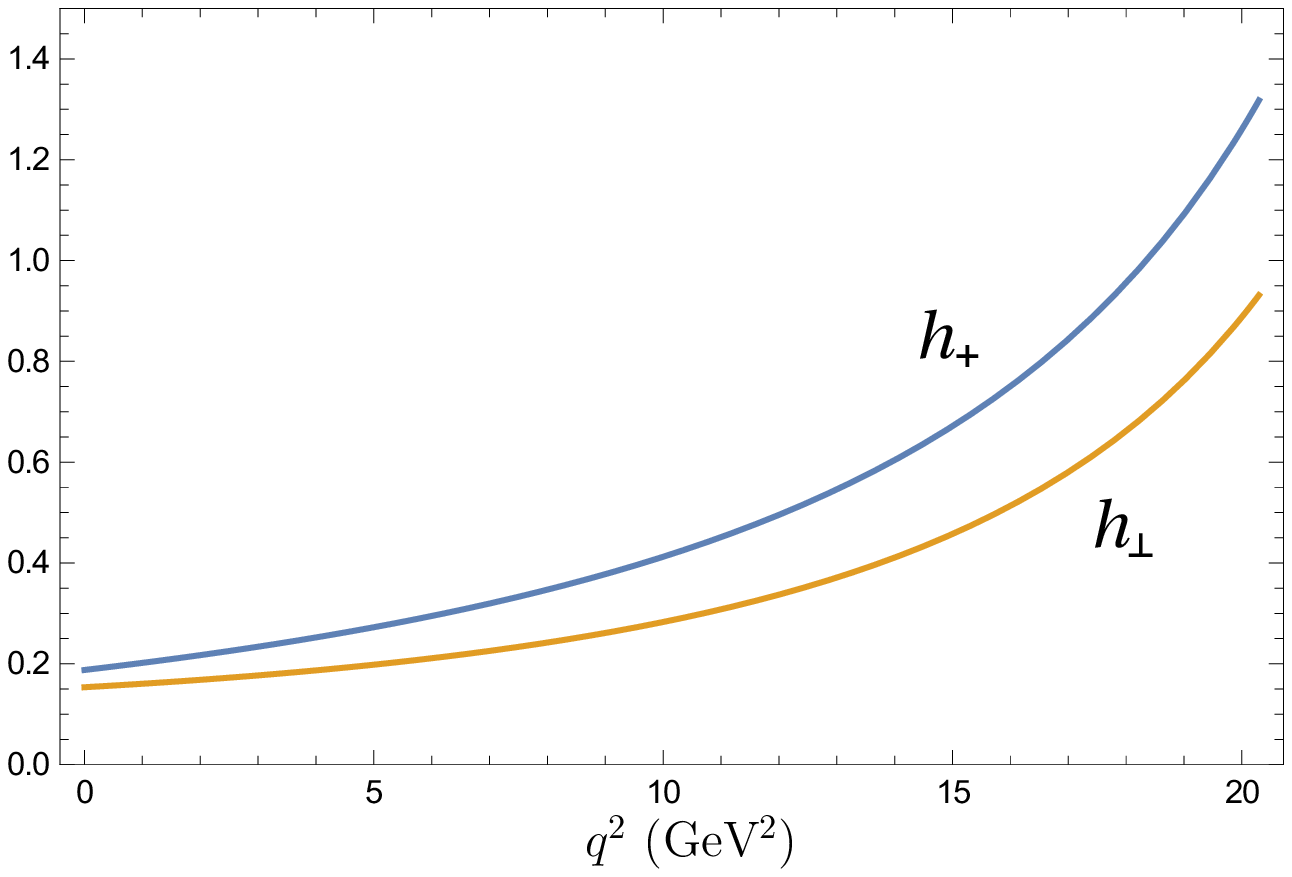}\ \
 \ \includegraphics[width=8cm]{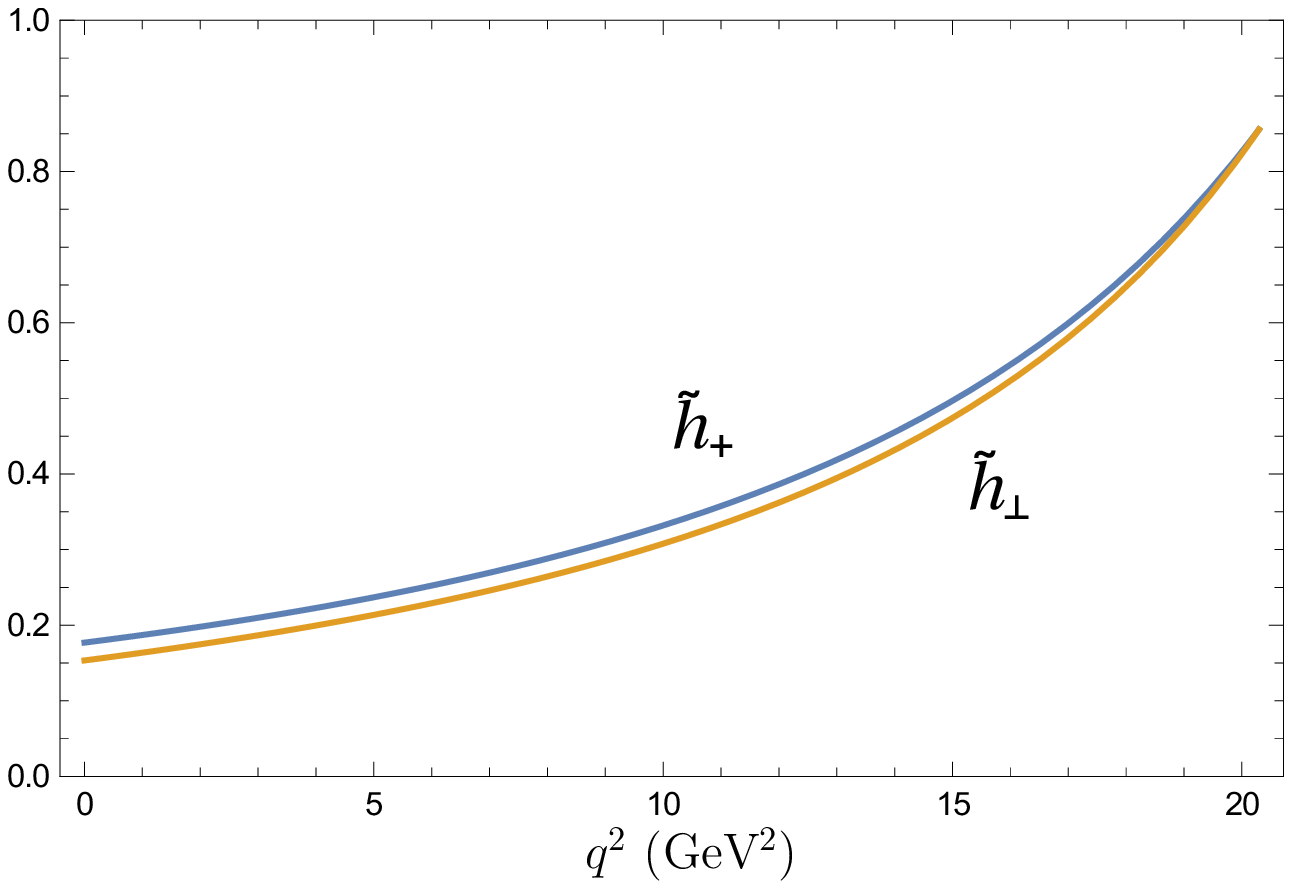}\\
\caption{Helicity form factors of the rare weak $\Lambda_b\to \Lambda$ transition.    } 
\label{fig:ffLbp}
\end{figure}

In Tables~\ref{compbpiff}, \ref{compff} we compare the calculated values
of the form factors with predictions of other approaches
\cite{gikls,aas,latt,ws}. The covariant constituent quark model was
employed in Ref.~\cite{gikls}, while form factors in Ref.~\cite{aas}
were calculated in the framework of the light-cone QCD sum rules. The
helicity form factors  (\ref{eq:hff}) were calculated using lattice
QCD with relativistic $b$ quarks in Ref.~\cite{latt}. QCD light-cone
sum rules with the account of next-to-leading perturbative corrections
were used in Ref.~\cite{ws}. Reasonable agreement between
substantially different approaches is observed. Note that most of the
previous theoretical methods determine the decay form factors in the
limited range of the momentum transfer squared $q^2$. Thus light-cone QCD
sum rules provide form factors near the maximum recoil point $q^2=0$,
while lattice calculations are performed for small values of the
recoil momentum near the point $q^2=q^2_{\rm max}$. Therefore, in such
approaches, the extrapolation of form factors to the whole kinematical
range is required using some phenomenological model prescriptions. The
important advantage of our model is the possibility to explicitly
determine the $q^2$ dependence of the decay form factors in the whole
kinematical range, which is rather broad, without extrapolations
and/or additional model assumptions.  In Fig.~\ref{fig:ffLbp} we plot
the helicity form factors calculated in our model.

\begin{sidewaystable}
\caption{Comparison of theoretical predictions for the form factors of
 the rare $\Lambda_b\to\Lambda$ transition at maximum
  recoil  $q^2=0$.  }
\label{compbpiff}
\begin{ruledtabular}
\begin{tabular}{ccccccccccc}
&$f^V_1(0)$&$f^V_2(0)$&$f^V_3(0)$&$f^A_1(0)$&$f^A_2(0)$&$f^A_3(0)$&$f^{TV}_1(0)$ & $f^{TV}_2(0)$& $f^{TA}_1(0)$ &
                                                                 $f^{TA}_2(0)$\\
\hline
this paper& 0.208& 0.032& 0.026& 0.125& $-0.003$& $-0.083$&$-0.029$ & $-0.153$ & 0.029 & $-0.153$\\
\cite{gikls}&0.107&0.043& $0.003$& 0.104& 0.003&$-0.052$&$-0.043$ & $-0.105$ & 0.003 & $-0.105$\\
\cite{aas} &0.322(112)& 0.011(4)& $-0.015(5)$ &0.318(110)&$0.013(4)$ &
                                                                       $-0.014(5)$& $-0.056(18)$ & $-0.295(105)$ & $0.101(35)$ & $-0.294(105)$ \\
\end{tabular}
\end{ruledtabular}

\vspace*{1cm}
\caption{Comparison of theoretical predictions for the helicity form factors of
the  rare $\Lambda_b\to\Lambda$ transition at zero recoil $q^2=q^2_{\rm
    max}$  and maximum recoil  $q^2=0$.  }
\label{compff}
\begin{ruledtabular}
\begin{tabular}{cccccccccccc}
form factor&&$f_+$&$f_\perp$&$f_0$&$g_+$&$g_\perp$&$g_0$&$h_+$ & $h_\perp$& $\tilde h_+$ &$\tilde h_\perp$\\
\hline
$f(q^2_{\rm max})$
&this paper& 1.12& 1.53& 0.96& 0.79& $0.79$& $1.07$&$1.32$ & $0.93$ & 0.86 & $0.86$\\
&\cite{latt}&1.37(9)&1.67(12)& $0.95(7)$& 0.91(6)& 0.91(6)&$1.37(9)$& $1.54(14)$ & $1.21(10)$ & $0.84(7)$ & $0.84(7)$\\

$f(0)$
&this paper& 0.208& 0.245& 0.207& 0.125& $0.128$& $0.125$& 0.188& 0.153& 0.177& 0.153\\
&\cite{latt}&0.212(35)&0.240(40)& $0.200(20)$& 0.197(90)& 0.169(70)&$0.199(20)$&0.228(40)&0.218(30)& $0.262(70)$& 0.235(60)\\
&\cite{ws} &0.18& 0.20& $0.18$ && & &0.21& 0.18\\
\end{tabular}
\end{ruledtabular}
\end{sidewaystable}

Now we can use the obtained form factors for the calculation of the
rare semileptonic $\Lambda_b\to\Lambda l^+l^-$ and rare radiative
$\Lambda_b\to\Lambda \gamma$ decay observables.

\section{Rare semileptonic $\Lambda_b$ baryon decays}
\label{sec:rd}

The effective Hamiltonian for the rare $b\to s$ transitions is given by \cite{bhi}       
\begin{equation}
  \label{eq:heff}
  {\cal H}_{\rm eff} =-\frac{4G_F}{\sqrt{2}}V_{ts}^*V_{tb}\sum_{i=1}^{10}c_i{\cal
      O}_i,
\end{equation}
where $G_F$ is the Fermi constant, $V_{tj}$ are
Cabibbo-Kobayashi-Maskawa matrix elements, $c_i$ are the Wilson coefficients
and ${\cal O}_i$ are the standard model  operators.

Then the matrix element of the $b\to s l^+l^-$ transition amplitude
between baryon states is given by
\begin{equation}
  \label{eq:mtl}
  {\cal M}(\Lambda_b\to \Lambda l^+l^-)=\frac{G_F\alpha}{2\sqrt{2}\pi}
  |V_{ts}^*V_{tb}|\left[T^{(1)}_\mu(\bar l\gamma^\mu l)+T^{(2)}_\mu
    (\bar l\gamma^\mu \gamma_5l)\right],
\end{equation}
where 
\begin{eqnarray}
  \label{eq:amp}
  T^{(1)}_\mu&=&c_9^{eff}\langle\Lambda|\bar s
                 \gamma^\mu(1-\gamma^5)b|\Lambda_b\rangle -\frac{2m_b}{q^2}c_7^{eff}\langle\Lambda|\bar s
                 i\sigma^{\mu\nu}q_\nu(1+\gamma^5)b|\Lambda_b\rangle,\cr
 T^{(2)}_\mu&=&c_{10}\langle\Lambda|\bar s
                 \gamma^\mu(1-\gamma^5)b|\Lambda_b\rangle
\end{eqnarray}
$T^{(m)}$ ($m=1,2$) are expressed through the form factors and the Wilson
coefficients. These amplitudes can be written in the helicity
basis $\varepsilon^{\mu}(\lambda)$ as follows
\begin{equation}
  \label{eq:hap}
  H^{m}_{\lambda_\Lambda,\lambda}=\varepsilon^{\dag \mu}(\lambda)T^{(m)}_\mu(\lambda_\Lambda),
\end{equation}
 where $\lambda_\Lambda$ is the
helicity of the final $\Lambda$ baryon and $\lambda=0,\pm 1,t$
correspond to longitudinal, transverse and time-like helicities,
respectively.

The helicity amplitudes for weak baryon transitions induced by vector ($V$) and axial
vector ($A$) currents are expressed in terms of the decay form factors
\cite{gikls}  in the following way
\begin{eqnarray}
  \label{eq:ha}
  H^{Vm,Am}_{+1/2,\, 0}&=&\sqrt{\frac{s_\mp}{q^2}}
\left[(M_{\Lambda_Q} \pm M_{\Lambda}){\cal F}^{Vm,Am}_1(q^2) \pm\frac{q^2}{M_{\Lambda_b}}
{\cal F}^{Vm,Am}_2(q^2)\right]\cr
 H^{Vm,Am}_{+1/2,\, 1}&=&\sqrt{2s_\mp}\left[
 {\cal F}^{Vm,Am}_1(q^2)\pm\frac{M_{\Lambda_Q} \pm M_{\Lambda}}{M_{\Lambda_b}}
{\cal F}^{Vm,Am}_2(q^2)\right],\cr
 H^{Vm,Am}_{+1/2,\,
                          t}&=&\sqrt{\frac{s_\pm}{q^2}}\left[(M_{\Lambda_Q}\mp M_{\Lambda}){\cal F}^{Vm,Am}_1(q^2) \pm\frac{q^2}{M_{\Lambda_b}}
{\cal F}^{Vm,Am}_2(q^2)\right],
\end{eqnarray}
where  the upper(lower)  sign corresponds  to $Vm(Am)$ and the
corresponding combinations of form factors are
\begin{eqnarray}
  \label{eq:ffe}
  {\cal
  F}^{V1,A1}_1(q^2)&=&c_9^{eff}f^{V,A}_1(q^2)\mp\frac{2m_b}{M_{\Lambda_b}}c_7^{eff}f^{TV,TA}_1(q^2),\cr
{\cal
                       F}^{V1,A1}_2(q^2)&=&c_9^{eff}f^{V,A}_2(q^2)\mp\frac{2m_bM_{\Lambda_b}}{q^2}c_7^{eff}f^{TV,TA}_2(q^2),\cr
{\cal
                                            F}^{V1,A1}_3(q^2)&=&c_9^{eff}f^{V,A}_3(q^2)+\frac{2m_b(M_{\Lambda_b}\mp M_\Lambda)}{q^2}c_7^{eff}f^{TV,TA}_1(q^2),
\end{eqnarray}
and
\begin{equation}
  \label{eq:ff2}
  {\cal F}^{V2,A2}_i(q^2)=c_{10} f^{V,A}_i(q^2), \qquad (i=1,2,3).
\end{equation}
The amplitudes for negative values of the helicities can be obtained
using the relation
$$H^{Vm,Am}_{-\lambda_\Lambda,\,-\lambda}=\pm H^{Vm,Am}_{\lambda_\Lambda,\, \lambda}.$$
The total helicity amplitude for the
$V-A$ current is then given by
$$H_{\lambda_\Lambda,\, \lambda}^m=H^{Vm}_{\lambda_\Lambda,\, \lambda}
-H^{Am}_{\lambda_\Lambda,\, \lambda}.$$

The values of the Wilson coefficients $c_i$ and of the effective
Wilson coefficient $c_7^{eff}$  are taken from Ref.~\cite{wc}. The 
effective  Wilson coefficient $ c_9^{\rm eff}$ contains additional
pertubative and long-distance contributions   
\begin{equation}
 \label{eq:ceff9}
  c_9^{\rm eff}=c_9+{\cal Y}_{\rm pert}(q^2)+{\cal Y}_{\rm BW}(q^2).
\end{equation}
The perturbative part is equal to
\begin{eqnarray}
  \label{eq:ypert}
{\cal Y}_{\rm pert}(q^2)&=&h\left(\frac{m_c}{m_b},\frac{q^2}{m_b^2}\right)(3c_1+c_2+3c_3+c_4+3c_5+c_6)\cr
&&-
\frac12 h\left(1,\frac{q^2}{m_b^2}\right)(4c_3+4c_4+3c_5+c_6)\cr
&&-\frac12
h\left(0,\frac{q^2}{m_b^2}\right)(c_3+3c_4)+\frac29(3c_3+c_4+3c_5+c_6),
\end{eqnarray}
where
\begin{eqnarray*} 
h\left(\frac{m_c}{m_b},  \frac{q^2}{m_b}\right) & = & 
- \frac{8}{9}\ln\frac{m_c}{m_b} +
\frac{8}{27} + \frac{4}{9} x 
-  \frac{2}{9} (2+x) |1-x|^{1/2} \left\{
\begin{array}{ll}
 \ln\left| \frac{\sqrt{1-x} + 1}{\sqrt{1-x} - 1}\right| - i\pi, &
 x \equiv \frac{4 m_c^2}{ q^2} < 1,  \\
 & \\
2 \arctan \frac{1}{\sqrt{x-1}}, & x \equiv \frac
{4 m_c^2}{ q^2} > 1,
\end{array}
\right. \\
h\left(0, \frac{q^2}{m_b} \right) & = & \frac{8}{27} - 
\frac{4}{9} \ln\frac{q^2}{m_b} + \frac{4}{9} i\pi.
\end{eqnarray*}
The long-distance (nonperturbative) contributions are assumed to originate from the $c\bar c$ resonances ($J/\psi, \psi'\dots$) and  have
a usual Breit-Wigner structure:
\begin{equation}
  \label{eq:ybw}
{\cal Y}_{\rm BW}(q^2)=\frac{3\pi}{\alpha^2} \sum_{V_i=J/\psi,\psi(2S)\dots}\frac{\Gamma(V_i\to l^+l^-)M_{V_i} }{M_{V_i}^2-q^2-iM_{V_i}\Gamma_{V_i}}.
\end{equation}
We include contributions of the vector $V_i(1^{--})$ charmonium states: $J/\psi$,
$\psi(2S)$, $\psi(3770)$, $\psi(4040)$, $\psi(4160)$ and $\psi(4415)$,
with their masses ($M_{V_i}$), leptonic [$\Gamma(V_i\to l^+l^-)$] and
total ($\Gamma_{V_i}$) decay widths taken from PDG \cite{pdg}.

The differential decay rate for the rare semileptonic $\Lambda_b$ baryon
decay to the $\Lambda$ baryon reads \cite{gikls}
\begin{equation}
  \label{eq:dgamma}
  \frac{d\Gamma(\Lambda_b\to \Lambda l^+l^-)}{dq^2}=\frac{G_F^2}{(2\pi)^3}
 \left(\frac{\alpha|V_{ts}^*V_{tb}|}{2\pi}\right)^2
  \frac{\lambda^{1/2}q^2}{48M_{\Lambda_b}^3}\sqrt{1-\frac{4m_l^2}{q^2}}{\cal H}_{tot},
\end{equation}
where $G_F$ is the Fermi constant, $V_{qQ}$ is the CKM matrix element, $\lambda\equiv
\lambda(M_{\Lambda_b}^2,M_{\Lambda}^2,q^2)=M_{\Lambda_b}^4+M_{\Lambda}^4+q^4-2(M_{\Lambda_b}^2M_{\Lambda}^2+M_{\Lambda}^2q^2+M_{\Lambda_b}^2q^2)$,
\begin{eqnarray}
  \label{eq:hh}
  {\cal H}_{tot}&=&\frac12({\cal H}_U^{11}+{\cal H}_U^{22}+{\cal H}_L^{11}+{\cal H}_L^{22}) \left(1-\frac{4m_l^2}{q^2}\right)+\frac{3m_l^2}{q^2}({\cal H}_U^{11}+{\cal H}_L^{11}+{\cal H}_S^{22}) ,\\
{\cal H}_U^{mm'}&=&\Re(H_{+1/2,+1}^mH_{+1/2,+1}^{\dag m'})+\Re(H_{-1/2,-1}^mH_{-1/2,-1}^{\dag m'}),\cr
{\cal H}_L^{mm'}&=&\Re(H_{+1/2,0}^mH_{+1/2,0}^{\dag m'})+\Re(H_{-1/2,0}^mH_{-1/2,0}^{\dag m'}),\cr
{\cal H}_S^{mm'}&=&\Re(H_{+1/2,t}^mH_{+1/2,t}^{\dag m'})+\Re(H_{-1/2,t}^mH_{-1/2,t}^{\dag m'}),\cr
{\cal H}_P^{mm'}&=&\Re(H_{+1/2,+1}^mH_{+1/2,+1}^{\dag m'})-\Re(H_{-1/2,-1}^mH_{-1/2,-1}^{\dag m'}),\cr
{\cal H}_{L_P}^{mm'}&=&\Re(H_{+1/2,0}^mH_{+1/2,0}^{\dag m'}-H_{-1/2,0}^mH_{-1/2,0}^{\dag m'}),\cr
{\cal H}_{S_P}^{mm'}&=&\Re(H_{+1/2,t}^mH_{+1/2,t}^{\dag m'}-H_{-1/2,t}^mH_{-1/2,t}^{\dag m'})\nonumber
\end{eqnarray}
and $m_l$ is the lepton mass.

The lepton angle differential decay distribution
is given by
\begin{equation}
  \label{eq:ddGl}
  \frac{d^2 \Gamma(\Lambda_b\to\Lambda l^+l^-)}{d q^2d\cos\theta}=\frac{d \Gamma(\Lambda_b\to\Lambda l^+l^-)}{d q^2}\left[\frac38(1+\cos^2\theta)(1-F_L)+ A_{FB}^\ell\cos\theta+\frac34F_L\sin^2\theta\right],\qquad
\end{equation}
where $\theta$ is the angle between the $\Lambda_b$ baryon and the positively
charged lepton in the dilepton rest frame.
The lepton forward-backward asymmetry is defined by \cite{gikls}
\begin{equation}
  \label{eq:afbl}
  A_{FB}^\ell(q^2)=\frac{\frac{d\Gamma}{dq^2}({\rm forward})-\frac{d\Gamma}{dq^2}({\rm backward})}{\frac{d\Gamma}{dq^2}}
=-\frac34\frac{\sqrt{1-\frac{4m_l^2}{q^2}}{\cal H}_P^{12}}{{\cal H}_{tot}}.\qquad
\end{equation}
The fraction of longitudinally polarized dileptons is expressed by 
\begin{equation}
F_L(q^2)=\frac{\frac12\left(1-\frac{4m_l^2}{q^2}\right)({\cal
    H}_L^{11}+{\cal H}_{L}^{22})+ \frac{m_l^2}{q^2}({\cal
    H}_{U}^{11}+{\cal H}_{L}^{11}+{\cal H}_{S}^{22})}{{\cal
    H}_{tot}}.\end{equation}

The hadron angle differential distribution of the decay
$\Lambda_b\to\Lambda(\to p\pi^-)l^+l^-$ is given by
\begin{equation}
  \label{eq:ddGh}
  \frac{d^2 \Gamma(\Lambda_b\to\Lambda l^+l^-)}{d
    q^2d\cos\theta_h}=Br(\Lambda\to p\pi^-)\frac{d \Gamma(\Lambda_b\to\Lambda l^+l^-)}{d q^2}\frac12\left(1+2A_{FB}^h\cos\theta_h\right),\qquad
\end{equation}
where $\theta_h$ is the angle between the proton and the $\Lambda$ baryon in
the $\Lambda_b$ rest frame. The hadron forward-backward asymmetry has
the form \cite{gikls}
\begin{equation}
  \label{eq:afbh}
  A_{FB}^h(q^2)
=\frac{\alpha_\Lambda}2\frac{\frac12\left(1-\frac{4m_l^2}{q^2}\right)({\cal
    H}_P^{11}+{\cal H}_P^{22}+{\cal H}_{L_P}^{11}+{\cal H}_{L_P}^{22})+\frac{3m_l^2}{q^2}({\cal
    H}_{P}^{11}+{\cal H}_{L_P}^{11}+{\cal H}_{S_P}^{22})}{{\cal H}_{tot}}.\qquad
\end{equation}

The other useful observable is the combined hadron-lepton
forward-backward asymmetry $A_{FB}^{h\ell}$. It is proportional to the
coefficient in front of the term $\cos\theta\cos\theta_h$ in the threefold
joint angular decay distribution for the decay of the unpolarized
$\Lambda_b$ \cite{gikls}. This asymmetry is expressed by
\begin{equation}
  \label{eq:afblh}
  A_{FB}^{h\ell}(q^2)
=-\frac34\frac{\alpha_\Lambda}2\frac{\sqrt{1-\frac{4m_l^2}{q^2}}{\cal H}_U^{12}}{{\cal H}_{tot}},\qquad
\end{equation}
where the value of the $\Lambda\to p\pi^-$ decay asymmetry
$\alpha_\Lambda$ is known from  experiment \cite{pdg}: $\alpha_\Lambda=0.642\pm0.013$.     

The average values of these quantities $\langle A_{FB}^\ell\rangle$,
$\langle A_{FB}^h\rangle$, $\langle A_{FB}^{h\ell}\rangle$ and
$\langle F_L\rangle$ should be calculated by separately integrating the numerators and denominators over $q^2$.

Substituting the form factors calculated in the previous
section into the expressions (\ref{eq:dgamma})--(\ref{eq:afblh}) we
calculate the rare $\Lambda_b$ decay branching fractions and asymmetry
parameters. 
We roughly estimate theoretical uncertainties of our results,
originating from the calculation of the decay form factors, to be
about 10\% (see discussion in Ref.~\cite{sllbdecay}.) 

\begin{figure}
  \centering
 \includegraphics[width=8cm]{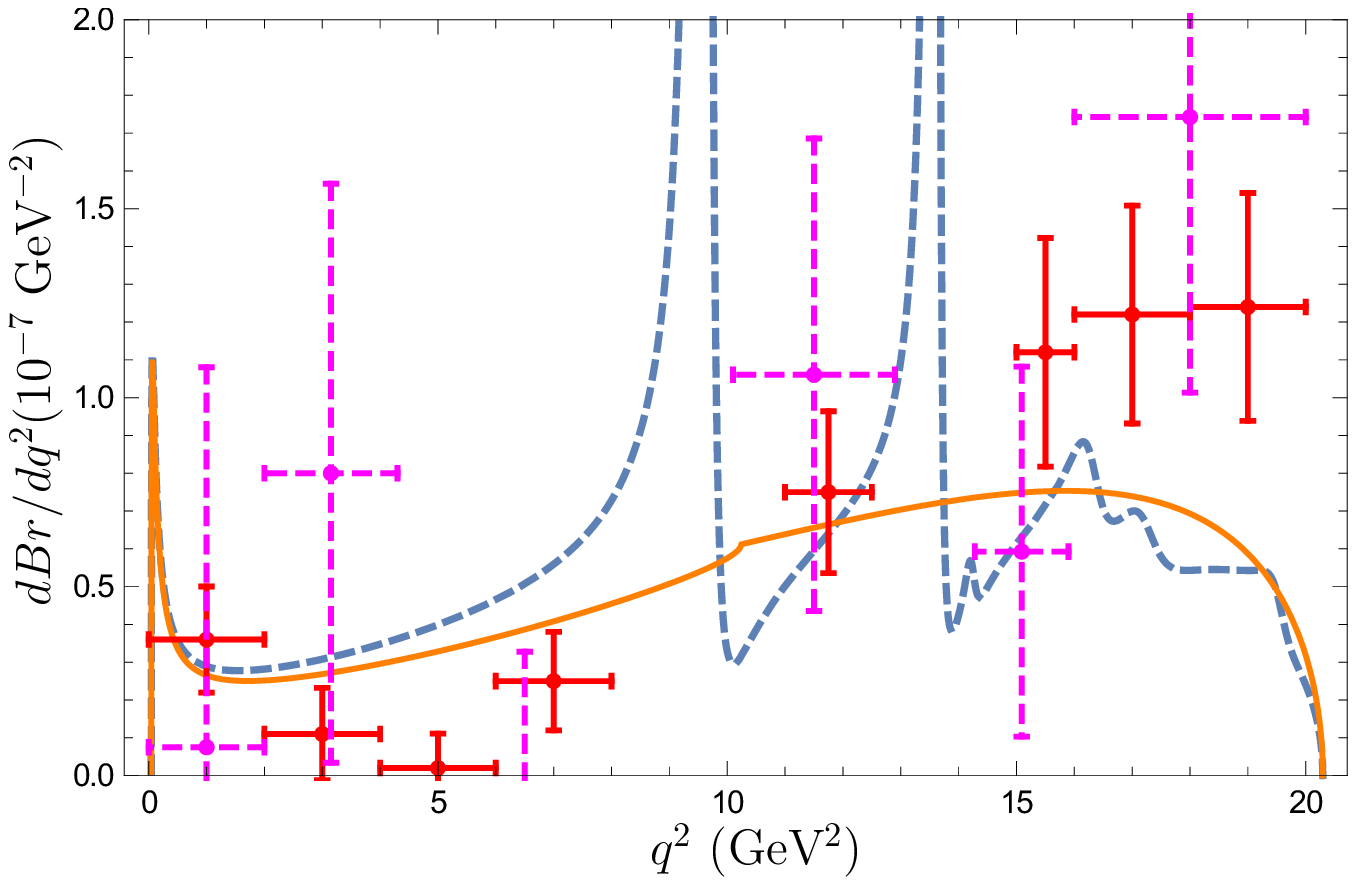}\ \
 \  \includegraphics[width=8cm]{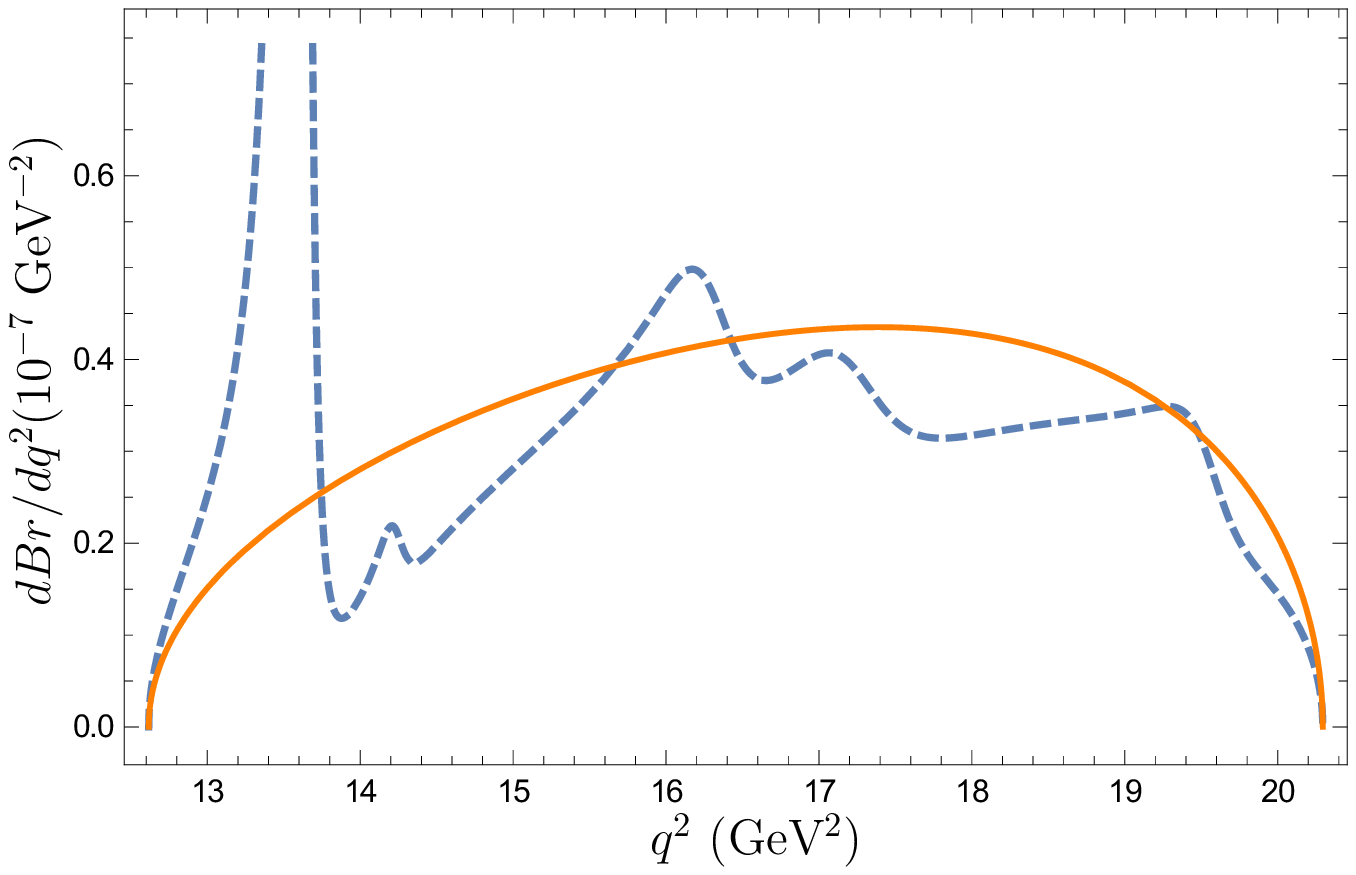}
  \caption{Predictions for the differential branching ratios for the
    $\Lambda_b\to \Lambda\mu^+\mu^-$ (left) and $\Lambda_b\to \Lambda\tau^+\tau^-$ (right)
    rare decays. Available experimental data from LHCb \cite{lhcb2015} are given
    by dots with solid error bars, CDF \cite{cdf} data are given by
    dots with dashed error bars. }
  \label{fig:brLb}
\end{figure}

\begin{figure}
  \centering
 \includegraphics[width=8cm]{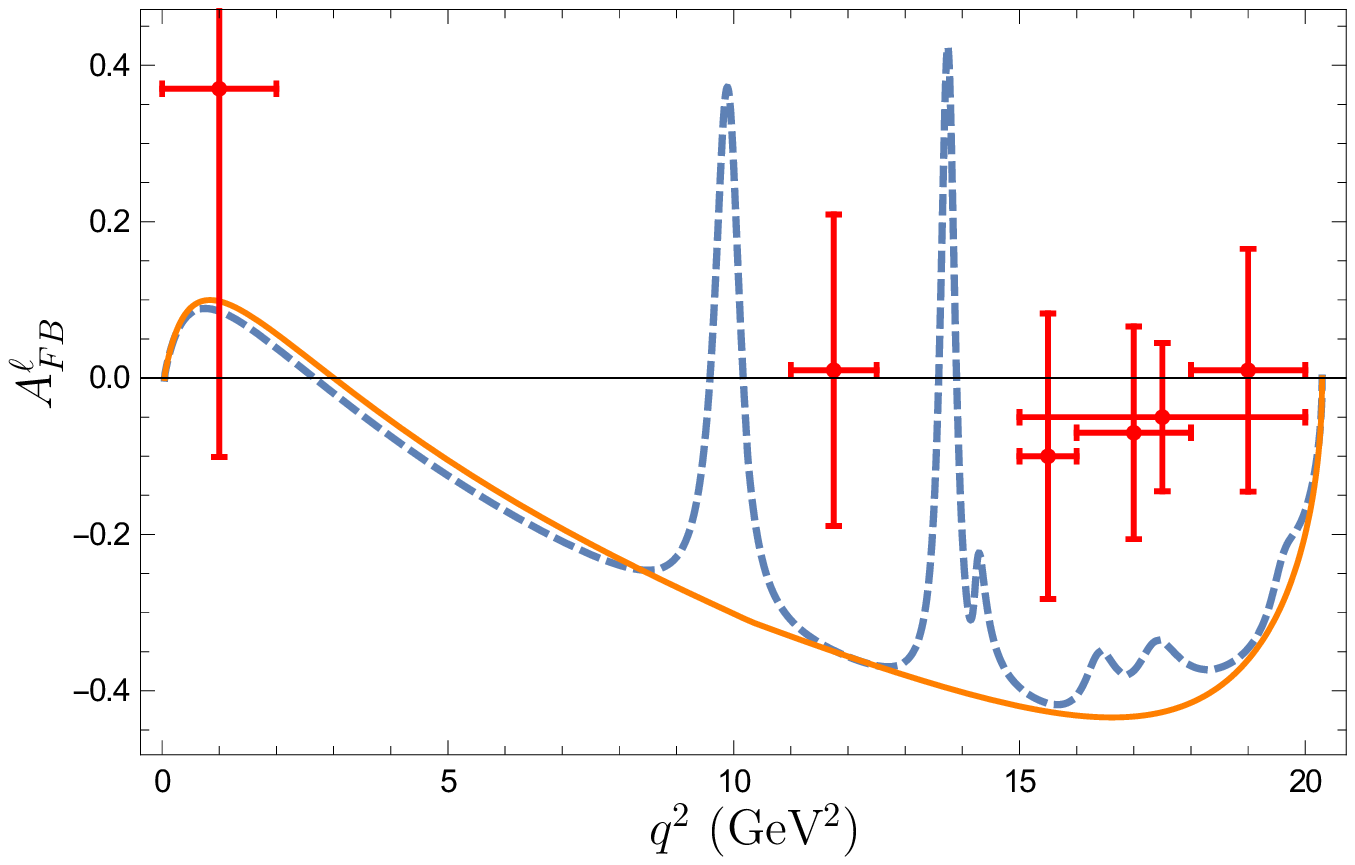}\ \
 \  \includegraphics[width=8cm]{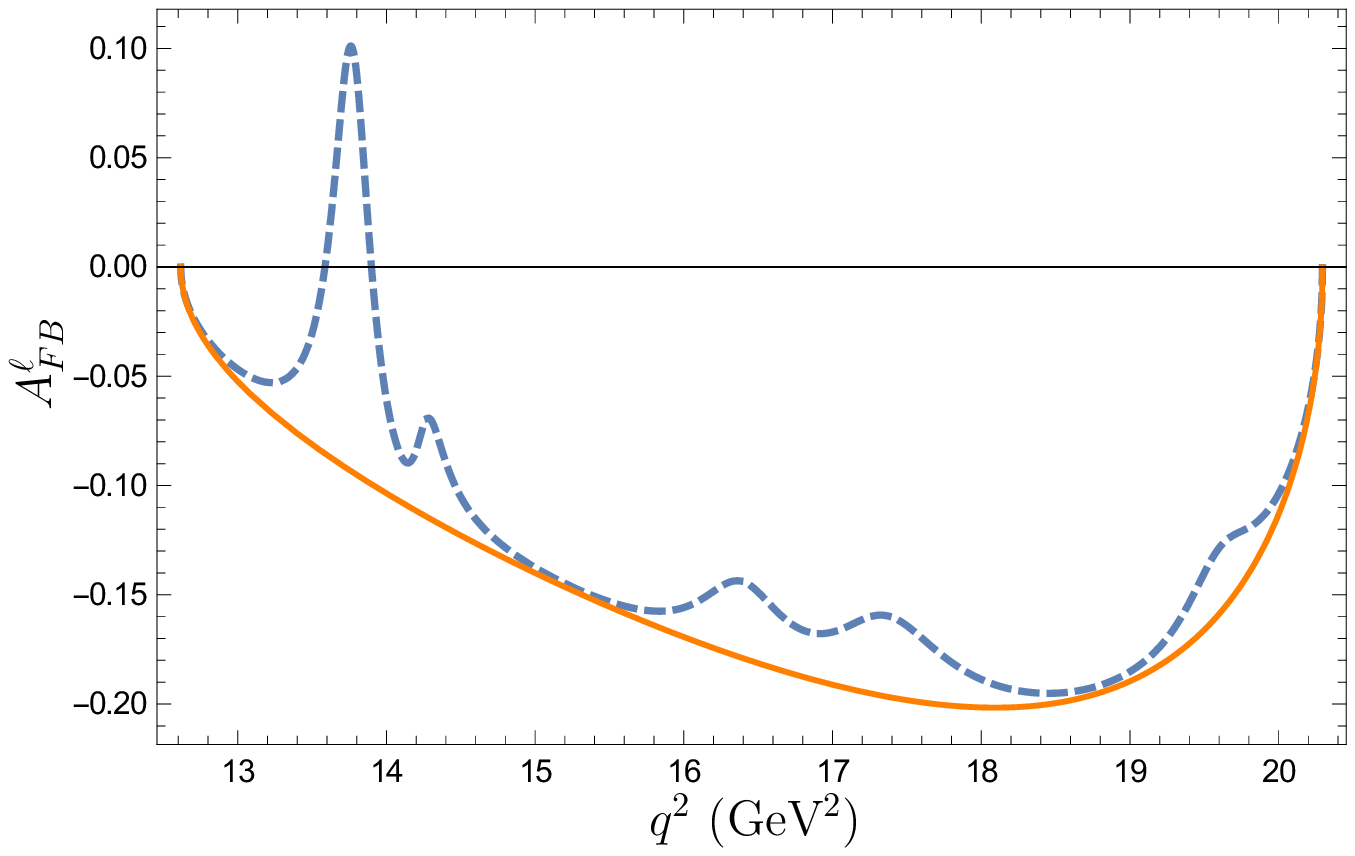}
\caption{Predictions for the lepton 
    forward-backward asymmetries  $A^\ell_{FB}(q^2)$ in the
    $\Lambda_b\to \Lambda \mu^+\mu^-$ (left) and $\Lambda_b\to \Lambda
    \tau^+\tau^-$ (right) rare decays. Data from LHCb \cite{lhcb2015} are given
    by dots with solid error bars.}
  \label{fig:afbl}
\end{figure}

\begin{figure}
  \centering
 \includegraphics[width=8cm]{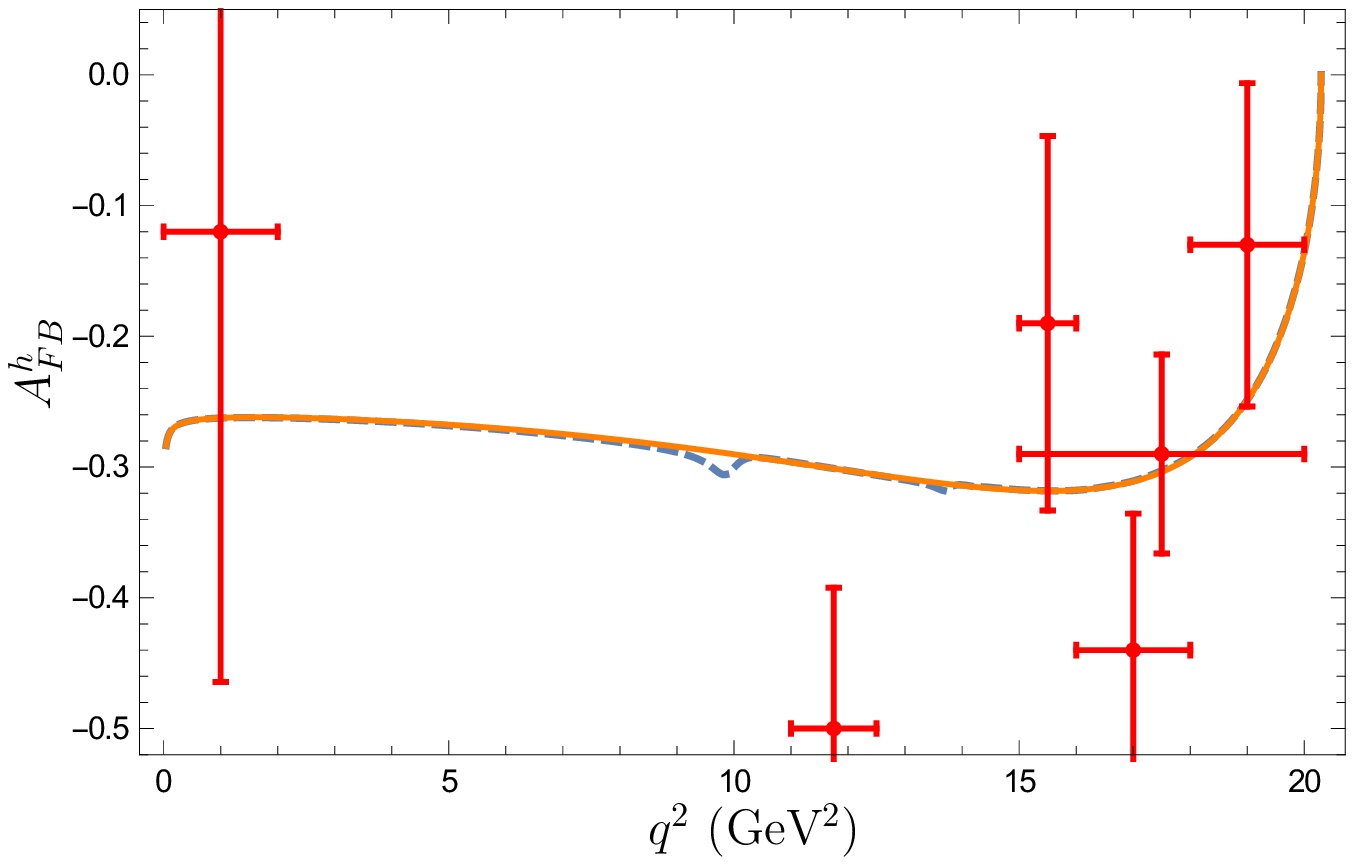}\ \
 \  \includegraphics[width=8cm]{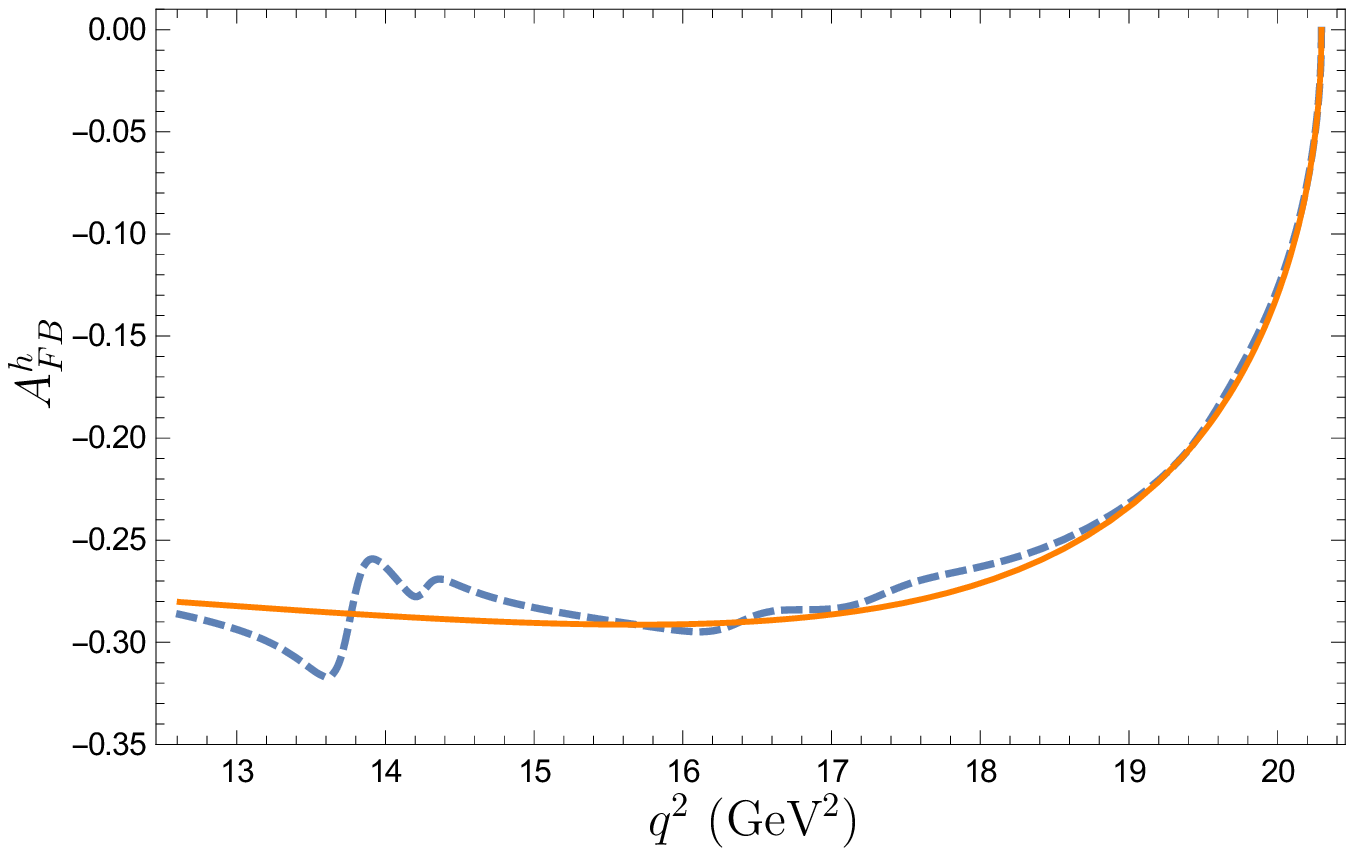}
  \caption{Predictions for the hadron forward-backward asymmetries
    $A^h_{FB}(q^2)$  in the  $\Lambda_b\to \Lambda \mu^+\mu^-$ (left) and $\Lambda_b\to \Lambda
    \tau^+\tau^-$ (right)  rare decays. Data from LHCb \cite{lhcb2015} are given
    by dots with solid error bars.}
  \label{fig:afbh}
\end{figure}

\begin{figure}
  \centering
 \includegraphics[width=8cm]{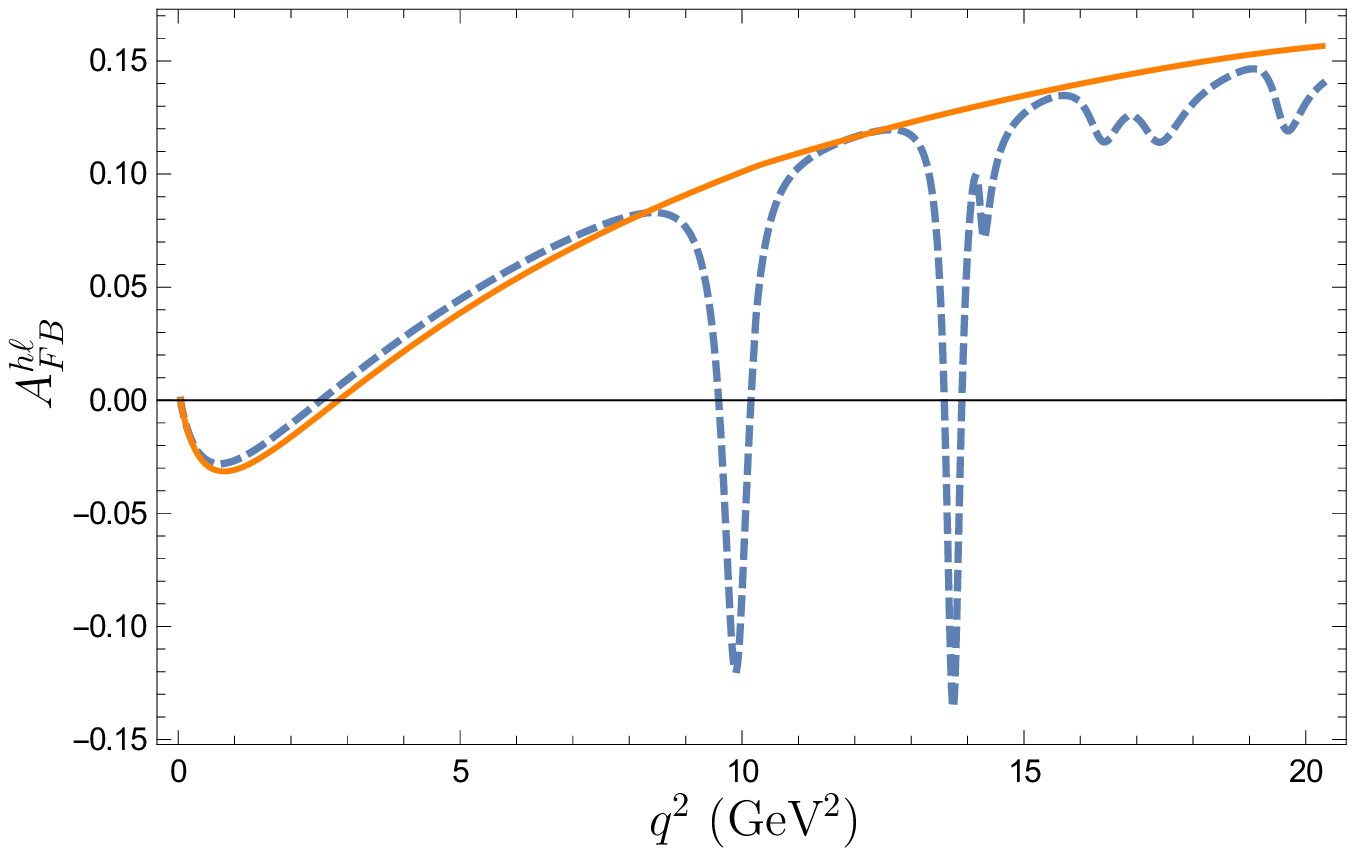}\ \
 \  \includegraphics[width=8cm]{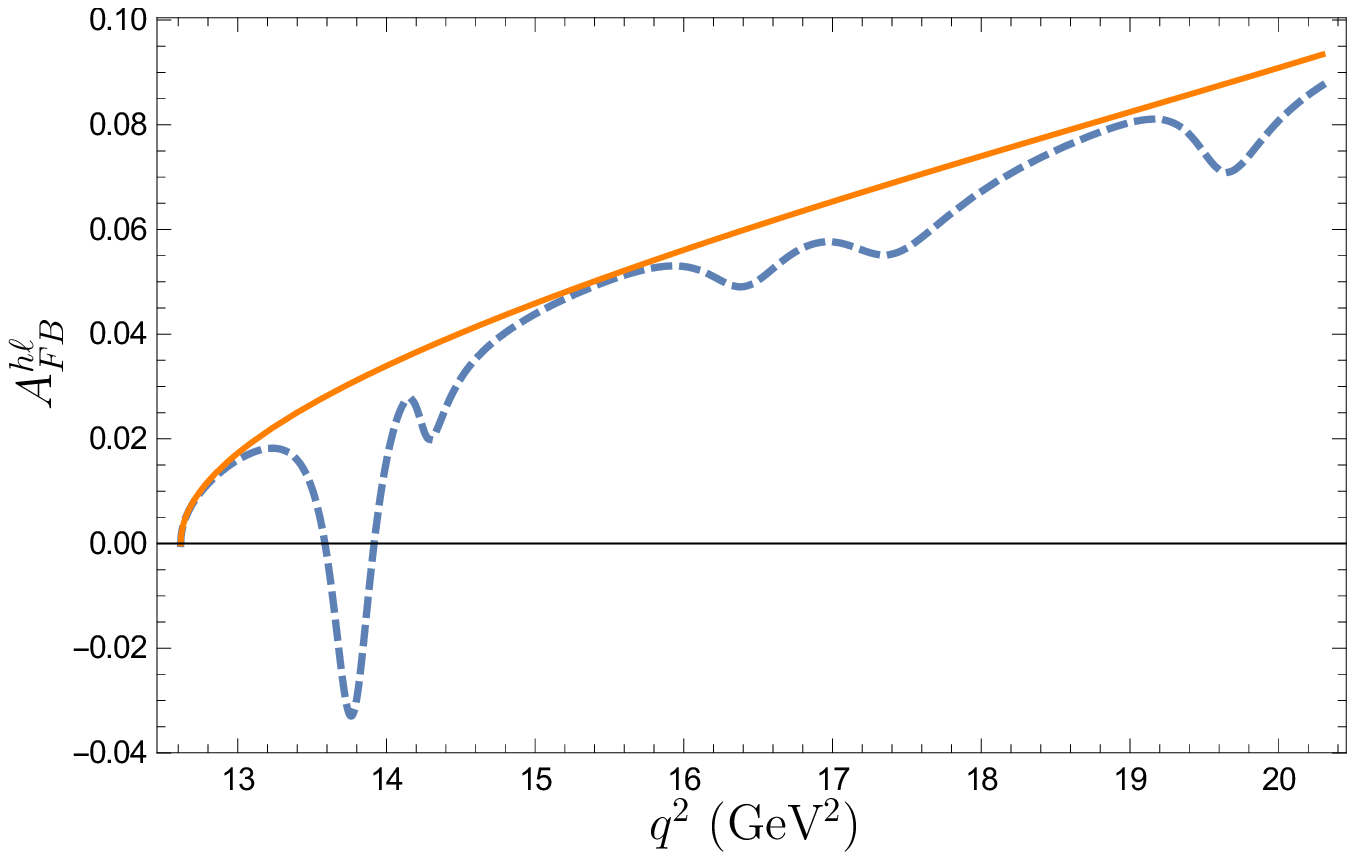}
  \caption{Predictions for the hadron-lepton forward-backward asymmetries
    $A^{h\ell}_{FB}(q^2)$   in the  $\Lambda_b\to \Lambda \mu^+\mu^-$ (left) and $\Lambda_b\to \Lambda
    \tau^+\tau^-$ (right) rare decays.}
  \label{fig:afbhl}
\end{figure}

\begin{figure}
  \centering
 \includegraphics[width=8cm]{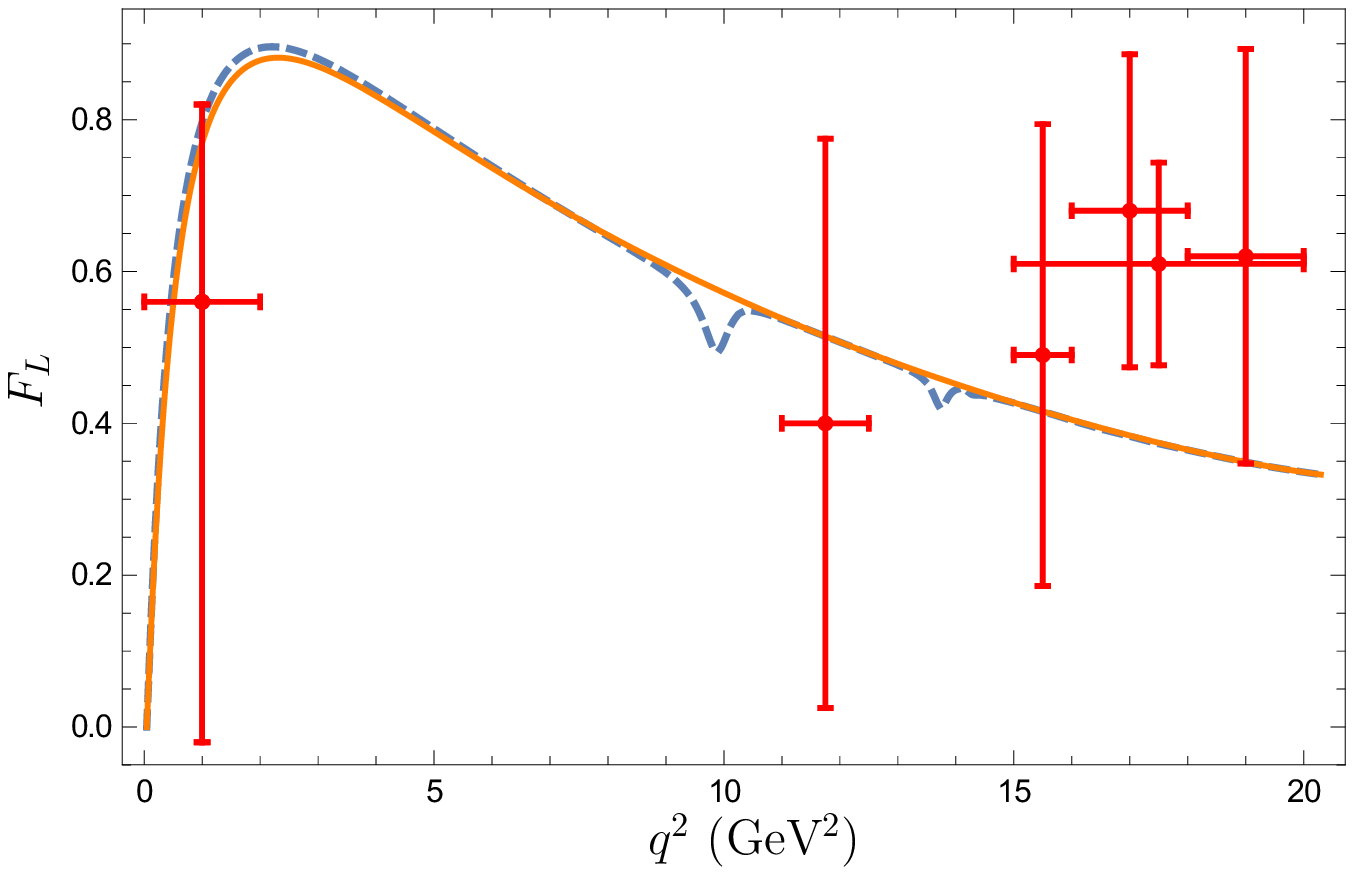}\ \
 \  \includegraphics[width=8cm]{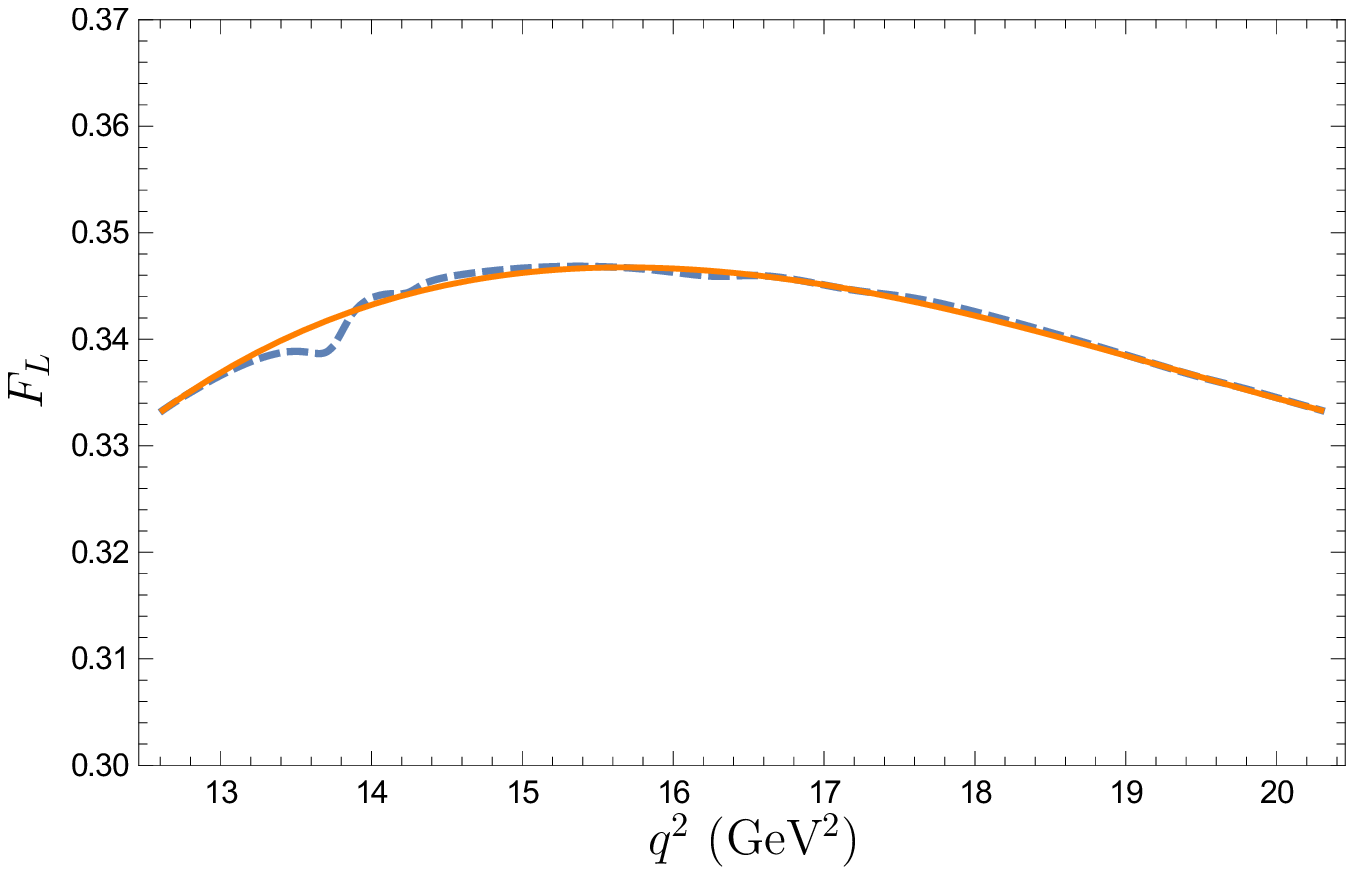}

  \caption{Prediction for the fraction of longitudinally polarized
    dileptons $F_L(q^2)$  in the  $\Lambda_b\to \Lambda \mu^+\mu^-$ (left) and $\Lambda_b\to \Lambda
    \tau^+\tau^-$ (right)
    rare decays. Data from LHCb \cite{lhcb2015} are given
    by dots with solid error bars. }
  \label{fig:cfl}
\end{figure}

In Figs.~\ref{fig:brLb}--\ref{fig:cfl} we plot our predictions for the
differential branching ratios $d Br/d q^2$, lepton $A^\ell_{FB}(q^2)$,
hadron $A^h_{FB}(q^2)$ and  hadron-lepton $A^{h\ell}_{FB}(q^2)$  forward-backward
asymmetries  as well as the fraction of longitudinally polarized
dileptons $F_L(q^2)$ for rare decays  $\Lambda_b\to \Lambda\mu^+\mu^-$
and $\Lambda_b\to \Lambda\tau^+\tau^-$ in comparison with available
experimental data \cite{lhcb2015,cdf}. By solid (dashed) lines we plot
theoretical results obtained without (with) inclusion of the
long-distance contributions to the Wilson coefficients coming from the charmonium
resonances. Experimental data for the  $\Lambda_b\to
\Lambda\mu^+\mu^-$ decay from the LHCb   \cite{lhcb2015} and CDF
\cite{cdf} Collaborations are plotted by dots with solid and dashed
error bars, respectively. 

In Table~\ref{brcomp} we compare different theoretical predictions \cite{gikls,aas,wll,llg,mr,glch}
for the total branching fractions of rare semileptonic $\Lambda_b$
decays with available experimental data \cite{pdg}. The presented values
include results of the relativistic and nonrelativistic quark
model calculations \cite{gikls,llg,mr} as well as evaluations based on
various versions of the light-cone QCD sum rules
\cite{aas,wll,glch}. At present, experimental data are available for
$\Lambda_b\to\Lambda\mu^+\mu^-$ decay only. The values obtained  in our
model and Refs.~\cite{gikls,llg,mr} agree well with data, while other
results are significantly larger. From Table~\ref{brcomp} we also see
that predictions for the rare $\Lambda_b\to \Lambda
\tau^+\tau^-$  decay vary significantly. Our results are close to those from
quark models
Refs.~\cite{gikls,mr}. On the other hand, the light-cone QCD sum rules
\cite{aas,wll} predict significantly  larger values while the
quark model \cite{llg}  gives a significantly lower value. Thus
experimental measurement of the rare $\Lambda_b\to \Lambda
\tau^+\tau^-$ decay branching fraction can help to discriminate
between theoretical approaches.   

The predicted values of the averaged asymmetries and
polarization fractions of the rare semileptonic $\Lambda_b$ decays are
given in Table~\ref{ass}. They include results obtained without (nonres.) and with (res.)
inclusion of the charmonium resonances in the
long-distance contribution (\ref{eq:ybw}) to the Wilson coefficient $c_7^{eff}$.
Table~\ref{comp15} contains a comparison of theoretical predictions for
the rare $\Lambda_b\to\Lambda \mu^+\mu^-$ decay observables  with the LHCb
experimental data \cite{lhcb2015} in the low recoil range 15~GeV$^2\le
q^2\le 20.28$~GeV$^2$. In this range the lattice results \cite{latt}
are the most reliable. The same interval of $q^2$ was chosen in Ref.~\cite{bfd} in
order to minimize the uncertainties from quark-hadron duality
violation. The authors of Ref.~\cite{bfd} combine predictions of the
lattice QCD for the rare decay form factors at low recoil with  form
factor relations in heavy quark effective theory. We see that
theoretical approaches give close values in this $q^2$ range. The
measured branching fraction $\langle Br\rangle$ and longitudinal
polarization  $\langle F_L\rangle$ are somewhat higher than
theoretical predictions, but agree with them within $2\sigma$.  The
measured hadron forward-backward asymmetry $\langle A^h_{FB}\rangle$
agrees well with predictions, while the experimental lepton forward-backward
asymmetry $\langle A^\ell_{FB}\rangle$ has significantly lower
absolute value.          

\begin{table}
\caption{Comparison of theoretical predictions for baryon rare decay
  branching fractions ($\times 10^{-6}$) with available experimental data. }
\label{brcomp}
\begin{ruledtabular}
\begin{tabular}{ccccccccc}
Decay& this paper & \cite{gikls} & \cite{aas}
  & \cite{wll}&\cite{llg}&\cite{mr}&\cite{glch} &Experiment \cite{pdg}\\
\hline
$\Lambda_b\to\Lambda e^+e^-$& 1.07& 1.0& 4.6(1.6)& &$1.21\sim 2.32$&
  &$2.03\left(^{26}_{9}\right)$\\
$\Lambda_b\to\Lambda \mu^+\mu^-$& 1.05& 1.0
                                     &4.0(1.2)&$6.1\left(^{5.8}_{1.7}\right)$&$0.53\sim
                                                                               0.89$&0.70&&$1.08(28)$\\
$\Lambda_b\to\Lambda \tau^+\tau^-$& 0.26& 0.2 &0.8(3)&$2.1\left(^{2.3}_{0.6}\right)$&$0.037\sim 0.083$&0.22&&\\
\end{tabular}
\end{ruledtabular}
\end{table}

\begin{table}
\caption{Predictions for the averaged rare decay asymmetries and polarization fractions. }
\label{ass}
\begin{ruledtabular}
\begin{tabular}{ccccccccc}

Decay& \multicolumn{2}{c}{$\langle A_{FB}^\ell\rangle$}&
                                                              \multicolumn{2}{c}{$\langle A_{FB}^h\rangle$}& \multicolumn{2}{c}{$\langle A_{FB}^{h\ell}\rangle$}& \multicolumn{2}{c}{$\langle F_L\rangle$}\\
& nonres.& res. &  nonres.& res. & nonres.& res. & nonres.& res. \\
\hline
$\Lambda_b\to \Lambda e^+e^-$&$-0.288$&$-0.294$&$-0.291$&$-0.299$&0.101&0.92&0.526&0.596\\
$\Lambda_b\to \Lambda \mu^+\mu^-$&$-0.286$&$-0.266$&$-0.288$&$-0.299$&0.101&0.92&0.525&0.544\\
$\Lambda_b\to \Lambda \tau^+\tau^-$&$-0.161$&$-0.127$&$-0.268$&$-0.313$&0.060&0.047&0.343&0.339\\
\end{tabular}
\end{ruledtabular}
\end{table}

\begin{table}
\caption{Comparison of theoretical predictions for the rare decay
  $\Lambda_b\to\Lambda \mu^+\mu^-$ observables with available
  experimental data in the range 15~GeV$^2\le q^2\le 20.28$~GeV$^2$. }
\label{comp15}
\begin{ruledtabular}
\begin{tabular}{cccccc}
Observable& nonres.& res. & \cite{latt} & \cite{bfd}&Experiment LHCb \cite{lhcb2015}\\
\hline
$\langle Br\rangle (\times 10^{-7})$& 3.54
             &3.22&3.99(37)&4.5(1.2)&5.9(1.4)\\
$\langle A^\ell_{\rm FB}\rangle$& $-0.40$
             &$-0.33$&$-0.350(13)$&$-0.29(5)$&$-0.05(9)$\\
$\langle A^h_{\rm FB}\rangle$& $-0.29$
             &$-0.29$&$-0.2710(92)$&$-0.26(3)$&$-0.29(8)$\\
$\langle A^{\ell h}_{\rm FB}\rangle$& $0.145$
             &$0.129$&$0.1398(43)$&$0.13(^2_3)$&\\
$\langle F_L\rangle$& $0.38$
             &$0.38$&$0.409(13)$&$0.4(1)$&$0.61(^{11}_{14})$\\
\end{tabular}
\end{ruledtabular}
\end{table}

In Tables~\ref{compBr}--\ref{compAhlFB} we present the comparison of our
predictions with lattice results \cite{latt} and experimental data
\cite{lhcb2015} for
the differential branching fractions $\langle Br/dq^2\rangle$,
forward-backward asymmetries $\langle A^\ell_{FB}\rangle$, $\langle
A^h_{FB}\rangle$, $\langle A^{hl}_{FB}\rangle$ and  longitudinal
polarization  $\langle F_L\rangle$ in several $q^2$ bins where such
data is available for the rare
$\Lambda_b\to\Lambda \mu^+\mu^-$ decay. 

\begin{table}
\caption{Comparison of theoretical predictions for the binned $\Lambda_b\to\Lambda \mu^+\mu^-$
 differential branching fractions $\langle Br/dq^2\rangle$ (in units
 of $10^{-7}$~GeV$^{-2}$) with available experimental data. }
\label{compBr}
\begin{ruledtabular}
\begin{tabular}{ccccc}
$q^2$ bin (GeV$^2$)& nonres. & res.  & \cite{latt} &Experiment LHCb \cite{lhcb2015}\\
\hline
$[0.1,2]$&0.31 &0.34&0.25(23)&0.36$(^{14}_{13})$\\
$[2,4]$&0.27 &0.31&0.18(12)&0.11$(^{12}_{9})$\\
$[4,6]$&0.33 &0.40&0.23(11)&0.02$(^{9}_{1})$\\
$[6,8]$&0.41 &0.57&0.307(94)&0.25$(^{13}_{12})$\\
$[11,12.5]$&0.66 &0.65&&0.75$(^{21}_{21})$\\
$[15,16]$&0.75 &0.72&0.796(75)&1.12$(^{30}_{30})$\\
$[16,18]$&0.73 &0.68&0.827(76)&1.22$(^{29}_{29})$\\
$[18,20]$&0.56 &0.49&0.665(68)&1.24$(^{30}_{30})$\\
$[1.1,6]$&0.29 &0.34&0.20(12)&0.09$(^{6}_{5})$\\
$[15,20]$&0.67&0.61&0.756(70)&1.20$(^{26}_{27})$\\
\end{tabular}
\end{ruledtabular}
\end{table}

\begin{table}
\caption{Comparison of theoretical predictions for the binned lepton
  forward-backward asymmetries $\langle A^\ell_{FB}\rangle$ in $\Lambda_b\to\Lambda \mu^+\mu^-$ decay with available experimental data. }
\label{compAlFB}
\begin{ruledtabular}
\begin{tabular}{ccccc}
$q^2$ bin (GeV$^2$)&  nonres. & res.  & \cite{latt} &Experiment LHCb \cite{lhcb2015}\\
\hline
$[0.1,2]$&0.078 &0.067&0.095(15)&0.37$(^{37}_{48})$\\
$[11,12.5]$&$-0.35$ &$-0.35$&&0.01$(^{20}_{19})$\\
$[15,16]$&$-0.42$ &$-0.41$&$-0.374(14)$&$-0.10(^{18}_{16})$\\
$[16,18]$&$-0.43$ &$-0.36$&$-0.372(13)$&$-0.07(^{14}_{13})$\\
$[18,20]$&$-0.35$ &$-0.32$&$-0.309(15)$&0.01$(^{16}_{15})$\\
$[15,20]$&$-0.40$&$-0.33$&$-0.350(13)$&$-0.05(^{10}_{10})$\\
\end{tabular}
\end{ruledtabular}
\end{table}

\begin{table}
\caption{Comparison of theoretical predictions for the binned hadron
  forward-backward asymmetries $\langle A^h_{FB}\rangle$ in $\Lambda_b\to\Lambda \mu^+\mu^-$ decay with available experimental data. }
\label{compAhFB}
\begin{ruledtabular}
\begin{tabular}{ccccc}
$q^2$ bin (GeV$^2$)&  nonres. &  res. & \cite{latt} &Experiment LHCb \cite{lhcb2015}\\
\hline
$[0.1,2]$&$-0.26$ &$-0.26$&$-0.310(18)$&$-0.12(^{34}_{32})$\\
$[11,12.5]$&$-0.30$ &$-0.30$&&$-0.50(^{11}_{4})$\\
$[15,16]$&$-0.32$ &$-0.32$&$-0.3069(83)$&$-0.19(^{14}_{16})$\\
$[16,18]$&$-0.31$ &$-0.31$&$-0.2891(90)$&$-0.44(^{10}_{6})$\\
$[18,20]$&$-0.25$ &$-0.25$&$-0.227(10)$&$-0.13(^{10}_{12})$\\
$[15,20]$&$-0.29$&$-0.29$&$-0.2710(92)$&$-0.29(^{8}_{8})$\\
\end{tabular}
\end{ruledtabular}
\end{table}

\begin{table}
\caption{Comparison of theoretical predictions for the binned
  longitudinal polarization  $\langle F_L\rangle$ in $\Lambda_b\to\Lambda \mu^+\mu^-$ decay with available experimental data. }
\label{compFL}
\begin{ruledtabular}
\begin{tabular}{ccccc}
$q^2$ bin (GeV$^2$)&  nonres. &  res. & \cite{latt} &Experiment LHCb \cite{lhcb2015}\\
\hline
$[0.1,2]$&$0.63$ &$0.66$&$0.465(84)$&$0.56(^{24}_{56})$\\
$[11,12.5]$&$0.51$ &$0.51$&&$0.40(^{37}_{36})$\\
$[15,16]$&$0.42$ &$0.41$&$0.454(20)$&$0.49(^{30}_{30})$\\
$[16,18]$&$0.38$ &$0.38$&$0.417(15)$&$0.68(^{15}_{21})$\\
$[18,20]$&$0.35$ &$0.35$&$0.3706(79)$&$0.62(^{24}_{27})$\\
$[15,20]$&$0.38$&$0.38$&$0.409(13)$&$0.61(^{11}_{14})$\\
\end{tabular}
\end{ruledtabular}
\end{table}

\begin{table}
\caption{Comparison of theoretical predictions for the binned hadron-lepton
  forward-backward asymmetries $\langle A^{h\ell}_{FB}\rangle$ in
  $\Lambda_b\to\Lambda \mu^+\mu^-$ decay. }
\label{compAhlFB}
\begin{ruledtabular}
\begin{tabular}{cccc}
$q^2$ bin (GeV$^2$)&  nonres. &  res. & \cite{latt} \\
\hline
$[0.1,2]$&$-0.024$ &$-0.021$&$-0.0302(51)$\\
$[2,4]$&$0.003$ &$0.010$&$-0.0169(99)$\\
$[4,6]$&$0.039$ &$0.045$&$0.021(13)$\\
$[6,8]$&$0.068$ &$0.072$&$0.053(13)$\\
$[15,20]$&$0.145$&$0.129$&$0.1398(43)$\\
\end{tabular}
\end{ruledtabular}
\end{table}

\section{Rare radiative $\Lambda_b$ baryon decay}
\label{sec:rrad}

The exclusive rare radiative decay rate $\Lambda_b\to \Lambda\gamma$ for
the emission of a real photon ($k^2=0$)  is given by
\begin{equation}
\label{rad}
\Gamma(\Lambda_b\to \Lambda\gamma)=
\frac{\alpha }{64\pi^4} G_F^2m_b^2M_{\Lambda_b}^3|V_{tb}V_{ts}|^2
|c_7^{\rm eff}(m_b)|^2 (|f_{2}^{TV}(0)|^2+|f_{2}^{TA}(0)|^2) 
\left(1-\frac{M_{\Lambda}^2}{M_{\Lambda_b}^2} \right)^3.
\end{equation}
Substituting the calculated values of the form factors
$f_{2}^{TV,TA}(0)$ we get the prediction for the branching fraction
which is given in Table~\ref{comprad}. In this table we also give
other theoretical values \cite{mw,gikls,wll,glch,cdfp} and the experimental
upper limit. Our result is consistent with the values from
Refs.~\cite{mw,wll}, but about a factor of 2 larger than the prediction
of the covariant constituent quark model \cite{gikls}. The result of
the light-cone QCD sum rule study \cite{glch} is about an order of
magnitude lower, while the value obtained within three-point QCD sum rules in the
heavy quark limit \cite{cdfp} is more than a factor 3 larger than other theoretical predictions. Thus the
measurement of the rare radiative $\Lambda_b\to \Lambda\gamma$ decay
branching fractions can discriminate between different approaches to
the form factor calculations.

\begin{table}
\caption{Comparison of theoretical predictions for the  rare radiative decay
 branching fraction  $Br(\Lambda_b\to\Lambda \gamma)$  ($\times 10^{-5}$) with available experimental data. }
\label{comprad}
\begin{ruledtabular}
\begin{tabular}{cccccccc}
Decay& this paper&\cite{mw} & \cite{gikls} 
  & \cite{wll}&\cite{glch}&\cite{cdfp} &Experiment \cite{pdg}\\
\hline
$\Lambda_b\to\Lambda \gamma$& 1.0& $0.77(^{22}_{19})$& 0.4 &$0.73(15)$
  &$0.061(^{14}_{13})$&3.1(6)&$<130$\\
\end{tabular}
\end{ruledtabular}
\end{table}

\section{Conclusions}
\label{sec:concl}

The form factors of the rare $\Lambda_b\to \Lambda$ baryon transitions were
obtained in the framework of the relativistic quark
model. Relativistic quark-diquark picture of baryons was employed. The
decay form factors are expressed through the overlap integrals of the
baryon wave functions. The obtained expressions take into account all
relativistic effects including the transformation of the baryon wave
functions from rest to the moving reference frame as well as
relativistic contributions of the intermediate negative energy
states. The momentum transfer squared  dependence of the form factors is
explicitly determined in the whole accessible kinematical range
without extrapolations and additional model assumptions. The
analytic expressions for the form factors which approximate
numerical results with high accuracy  are given in
Eq.~(\ref{fitff}). Such an approach
significantly improves the accuracy of theoretical predictions since  most
of the previous calculations determine form factors in a single
kinematical point or in the limited kinematical range and then require
assumptions on the form factor $q^2$ dependence or extrapolations to the whole
$q^2$ range.  

The branching fractions, various forward-backward asymmetries and
polarization fractions for the rare semileptonic $\Lambda_b\to\Lambda
l^+l^-$ decays were calculated in the framework of the standard model
using the obtained form factors. Calculations were performed both with and
without long-distance contributions to the effective Wilson 
coefficient  $c_7^{eff}$ arising from the account  of
resonances corresponding to charmonium states. Detailed comparison of
the obtained predictions with previous quark model, light-cone QCD sum
rules and lattice calculations as well as  experimental data
for the decay $\Lambda_b\to\Lambda\mu^+\mu^-$ is given. The calculations
of the decay observables are performed for several $q^2$ bins for
which experimental data is available \cite{lhcb2015}. Good agreement
of our predictions with lattice \cite{latt} results is found. In
general reasonable agreement of the calculated and measured observables of the  $\Lambda_b\to\Lambda\mu^+\mu^-$ decay is achieved,
however in some $q^2$ bins deviations are found to be about
$2\sigma$. Therefore additional and more precise measurements are needed for
confirming the standard model predictions or revealing possible
deviations from them in rare baryon decays.  

\acknowledgements
We are grateful to A. Dolgolenko, D. Ebert, M. Ivanov, J. K\"orner  and V. Matveev for valuable  discussions and support.

\appendix*
\section{Form factors of rear the $\Lambda_b\to\Lambda$ transitions}

The expressions for vector and axial vector decay form factors are
given in Ref.~\cite{sllbdecay}. The tensor and pseudo tensor decay form factors are as follows (the value
of the long range anomalous chromomagnetic quark moment $\kappa=-1$).

\subsection{Tensor form factors}
\label{sec:1}
\begin{equation}
f_1^{TV}(q^2)=f_1^{TV(1)}(q^2)+\varepsilon
f_1^{TV(2)S}(q^2)+(1-\varepsilon)f_1^{TV(2)V}(q^2);
\end{equation}
\begin{eqnarray}
\label{eq:f1TV}
f_1^{TV(1)}(q^2)&=&-\int \frac{d^3p}{(2\pi)^3} \bar\Psi_F\left({\bf
    p}+\frac{2\epsilon_d}{E_F+M_F}{\bf
    \Delta}\right)\sqrt{\frac{\epsilon_Q(p)+m_Q}{2\epsilon_Q(p)}}\sqrt{\frac{\epsilon_q(p+\Delta)+m_q}{2\epsilon_q(p+\Delta)}}\cr
&&\times
\Biggl\{\frac{\epsilon_d}{E_F+M_F}\Biggl[\frac{M_F}{\epsilon_q(p+\Delta)+m_q}+\frac{M_I}{\epsilon_Q(p)+m_Q}\cr
&&+\frac{(M_I+M_F)\epsilon_d}{(\epsilon_q(p+\Delta)+m_q)(\epsilon_Q(p)+m_Q)}\frac{E_F-M_F}{E_F+M_F}\Biggr]\cr
&&+ \frac{\bf p
  \Delta}{{\bf \Delta}^2}\Biggl[\frac{M_F}{\epsilon_q(p+\Delta)+m_q}
-\frac{M_I}{\epsilon_Q(p)+m_Q}\Biggr]\cr
&&-\frac13\frac{M_I+M_F}{E_F+M_F}\frac{{\bf
    p}^2}{(\epsilon_q(p+\Delta)+m_q)(\epsilon_Q(p)+m_Q)}\Biggr\}\Psi_I({\bf
  p}); 
\end{eqnarray}
\begin{eqnarray}
  \label{eq:f1TVs}
f_1^{TV(2)S}(q^2)&=&\int \frac{d^3p}{(2\pi)^3} \bar\Psi_F\left({\bf
    p}+\frac{2\epsilon_d}{E_F+M_F}{\bf
    \Delta}\right)\sqrt{\frac{\epsilon_Q(p)+m_Q}{2\epsilon_Q(p)}}\sqrt{\frac{\epsilon_q(\Delta)+m_q}{2\epsilon_q(\Delta)}}\cr
&&\times\Biggl\{\frac{1}{2\epsilon_Q(\Delta)(\epsilon_Q(\Delta)+m_Q)}\frac{\bf p
  \Delta}{{\bf\Delta}^2}M_I\left[M_I-\epsilon_Q(p)-\epsilon_d(p)\right]\cr
&&
-\frac{1}{2\epsilon_q(\Delta)(\epsilon_q(\Delta)+m_q)}\Biggl[\epsilon_q(\Delta)-m_q+(E_F-M_F)\left(1-\frac{\epsilon_d}{E_F+M_F}\right)+\frac{\bf p
  \Delta}{{\bf\Delta}^2}E_F\Biggr]\cr
&&\Biggl[M_F-\epsilon_q\left({\bf
    p}+\frac{2\epsilon_d}{E_F+M_F}{\bf
    \Delta}\right) -\epsilon_d\left({\bf  p}+\frac{2\epsilon_d}{E_F+M_F}{\bf\Delta}\right)\Biggr]\Biggr\}\Psi_I({\bf p});
\end{eqnarray}
\begin{eqnarray}
  \label{eq:f1TVv}
f_1^{TV(2)V}(q^2)&=&\int \frac{d^3p}{(2\pi)^3} \bar\Psi_F\left({\bf
    p}+\frac{2\epsilon_d}{E_F+M_F}{\bf
    \Delta}\right)\sqrt{\frac{\epsilon_Q(p)+m_Q}{2\epsilon_Q(p)}}\sqrt{\frac{\epsilon_q(\Delta)+m_q}{2\epsilon_q(\Delta)}}\cr
&&\times\Biggl\{\frac{1}{2\epsilon_Q(\Delta)(\epsilon_Q(\Delta)+m_Q)}\Biggl(\left[\epsilon_Q(\Delta)-m_Q+(E_F-M_F)\left(1-\frac{\epsilon_d}{E_F+M_F}\right)\right]\cr
&&\times\left[\frac{\bf p
  \Delta}{{\bf\Delta}^2}\frac{M_I}{2E_d}-\frac{\epsilon_d}{2m_Q}\left(\frac{M_I+M_F}{E_F+M_F}\right)\left(1-\frac{E_F-M_F}{M_I+M_F}\right)\right]+\frac{\bf p
  \Delta}{{\bf\Delta}^2}M_I\Biggr)\cr
&&\times\left[M_I-\epsilon_Q(p)-\epsilon_d(p)\right]
-\frac{1}{2\epsilon_q(\Delta)(\epsilon_q(\Delta)+m_q)}\Biggl(\Biggl[\epsilon_q(\Delta)-m_q\cr
&&+(E_F-M_F)\left(1+\frac{\epsilon_d}{E_F+M_F}\right)\Biggr]\left[1-\frac{\epsilon_d}{m_q}\frac{M_F}{E_F+M_F}-\frac{\bf p
  \Delta}{{\bf\Delta}^2}\frac{E_F}{2E_d}\right]+\frac{\bf p
  \Delta}{{\bf\Delta}^2}E_F\Biggr)\cr
&&\times\Biggl[M_F-\epsilon_q\left({\bf
    p}+\frac{2\epsilon_d}{E_F+M_F}{\bf
    \Delta}\right) -\epsilon_d\left({\bf  p}+\frac{2\epsilon_d}{E_F+M_F}{\bf\Delta}\right)\Biggr]\Biggr\}\Psi_I({\bf p});
\end{eqnarray}

\begin{equation}
f_2^{TV}(q^2)=f_2^{TV(1)}(q^2)+\varepsilon
f_2^{TV(2)S}(q^2)+(1-\varepsilon)f_2^{TV(2)V}(q^2);
\end{equation}

\begin{eqnarray}
\label{eq:fTV2}
f_2^{TV(1)}(q^2)&=&-\int \frac{d^3p}{(2\pi)^3} \bar\Psi_F\left({\bf
    p}+\frac{2\epsilon_d}{E_F+M_F}{\bf
    \Delta}\right)\sqrt{\frac{\epsilon_Q(p)+m_Q}{2\epsilon_Q(p)}}\sqrt{\frac{\epsilon_q(p+\Delta)+m_q}{2\epsilon_q(p+\Delta)}}\cr
&&\times\Biggl\{1+\frac{\epsilon_d}{E_F+M_F}\Biggl[\frac{M_I-M_F}{M_I}\left(\frac{M_F}{\epsilon_q(p+\Delta)+m_q}-\frac{M_I}{\epsilon_Q(p)+m_Q}\right)\cr
&&
+(E_F-M_F)\left(\frac1{\epsilon_q(p+\Delta)+m_q}+\frac{1}{\epsilon_Q(p)+m_Q}\right)\cr
&&
-\frac{\epsilon_d(E_F+M_F)}{(\epsilon_q(p+\Delta)+m_q)(\epsilon_Q(p)+m_Q)}\left(
   \frac{M_I^2+M_F^2}{M_I(E_F+M_F)}-
   \frac{E_F-M_F}{E_F+M_F}\right)\Biggr]\cr
&&+\frac{\bf
  p \Delta}{{\bf
   \Delta}^2}\Biggl[\frac{M_I-M_F}{M_I}\left(\frac{M_F}{\epsilon_q(p+\Delta)+m_q}+\frac{M_I}{\epsilon_Q(p)+m_Q}\right)\cr
&&+(E_F-M_F)\left(\frac1{\epsilon_q(p+\Delta)+m_q}-\frac{1}{\epsilon_Q(p)+m_Q}\right)\Biggr]\cr
&& -\frac{{\bf
    p}^2}{(\epsilon_q(p+\Delta)+m_q)(\epsilon_Q(p)+m_Q)}\left[1-\frac13\frac{(M_I+M_F)^2}{M_I(E_F+M_F)}\right]
\Biggr\}\Psi_I({\bf p});
\end{eqnarray}

\begin{eqnarray}
  \label{eq:f2TVs}
f_2^{TV(2)S}(q^2)&=&\int \frac{d^3p}{(2\pi)^3} \bar\Psi_F\left({\bf
    p}+\frac{2\epsilon_d}{E_F+M_F}{\bf
    \Delta}\right)\sqrt{\frac{\epsilon_Q(p)+m_Q}{2\epsilon_Q(p)}}\sqrt{\frac{\epsilon_q(\Delta)+m_q}{2\epsilon_q(\Delta)}}\cr
&&\times\Biggl\{\frac{1}{2\epsilon_Q(\Delta)(\epsilon_Q(\Delta)+m_Q)}\Biggl(\left[\epsilon_Q(\Delta)-m_Q+(E_F-M_F)\left(1-\frac{\epsilon_d}{E_F+M_F}\right)\right]\cr
&&+\frac{\bf p
  \Delta}{{\bf\Delta}^2}(E_F+M_F)\left(1-\frac{M_I+M_F}{E_F+M_F}\right)\Biggr)\left[M_I-\epsilon_Q(p)-\epsilon_d(p)\right]\cr
&&
+\frac{1}{2\epsilon_q(\Delta)(\epsilon_q(\Delta)+m_q)}\Biggl[\frac{M_F}{M_I}\left(\epsilon_q(\Delta)-m_q+(E_F-M_F)\left(1-\frac{\epsilon_d}{E_F+M_F}\right)\right)\cr
&&-\frac{\bf p
  \Delta}{{\bf\Delta}^2}(E_F+M_F)\left(1-\frac{E_F}{M_I}\frac{M_I+M_F}{E_F+M_F}\right)\Biggr]\cr
&&\times\Biggl[M_F-\epsilon_q\left({\bf
    p}+\frac{2\epsilon_d}{E_F+M_F}{\bf
    \Delta}\right) -\epsilon_d\left({\bf  p}+\frac{2\epsilon_d}{E_F+M_F}{\bf\Delta}\right)\Biggl]\Psi_I({\bf p});
\end{eqnarray}

\begin{eqnarray}
  \label{eq:f2TVv}
f_2^{TV(2)V}(q^2)&=&\int \frac{d^3p}{(2\pi)^3} \bar\Psi_F\left({\bf
    p}+\frac{2\epsilon_d}{E_F+M_F}{\bf
    \Delta}\right)\sqrt{\frac{\epsilon_Q(p)+m_Q}{2\epsilon_Q(p)}}\sqrt{\frac{\epsilon_q(\Delta)+m_q}{2\epsilon_q(\Delta)}}\cr
&&\times\Biggl\{\frac{1}{2\epsilon_Q(\Delta)(\epsilon_Q(\Delta)+m_Q)}\Biggl(\left[\epsilon_Q(\Delta)-m_Q+(E_F-M_F)\left(1-\frac{\epsilon_d}{E_F+M_F}\right)\right]\cr
&&\times\Biggl[1-\frac{\epsilon_d}{m_Q}\left(1-\frac{(M_I+M_F)^2}{2M_I(E_F+M_F)}\left(1-\frac{E_F-M_F}{M_I+M_F}\right)\right)\cr
&&-\frac{\bf p \Delta}{{\bf\Delta}^2}\frac{E_F+M_F}{2E_d}\left(1-\frac{M_I+M_F}{E_F+M_F}-\frac{\epsilon_d}{m_Q}\frac{E_F-M_F}{E_F+M_F}\right) \Biggr]
\cr
&&+\frac{\bf p
  \Delta}{{\bf\Delta}^2}(E_F+M_F)\left(1-\frac{M_I+M_F}{E_F+M_F}-\frac{\epsilon_d}{m_Q}\frac{E_F-M_F}{E_F+M_F}\right)\Biggr)\left[M_I-\epsilon_Q(p)-\epsilon_d(p)\right]\cr
&&
+\frac{1}{2\epsilon_q(\Delta)(\epsilon_q(\Delta)+m_q)}\Biggl[\frac{M_F}{M_I}\left(\epsilon_q(\Delta)-m_q+(E_F-M_F)\left(1-\frac{\epsilon_d}{E_F+M_F}\right)\right)
\cr
&&\times\Biggl(1-\frac{\epsilon_d}{m_q}\left(\frac{M_I+M_F}{E_F+M_F}-\frac{M_I}{M_F}\right)-\frac{\bf
   p \Delta}{{\bf\Delta}^2}\frac{M_I}{M_F}\frac{E_F+M_F}{2E_d}
\cr
&&\times\left(1-\frac{E_F}{M_I}\frac{M_I+M_F}{E_F+M_F}-\frac{\epsilon_d}{m_q}\frac{E_F-M_F}{E_F+M_F}\right) \Biggr)
\cr
&&-\frac{\bf p
  \Delta}{{\bf\Delta}^2}(E_F+M_F)\left(1-\frac{E_F}{M_I}\frac{M_I+M_F}{E_F+M_F}-\frac{\epsilon_d}{m_q}\frac{E_F-M_F}{E_F+M_F}\right)\Biggr]\cr
&&\times\Biggl[M_F-\epsilon_q\left({\bf
    p}+\frac{2\epsilon_d}{E_F+M_F}{\bf
    \Delta}\right) -\epsilon_d\left({\bf  p}+\frac{2\epsilon_d}{E_F+M_F}{\bf\Delta}\right)\Biggl]\Psi_I({\bf p});
\end{eqnarray}

\subsection{Pseudo tensor form factors}
\label{sec:2}
\begin{equation}
f_1^{TA}(q^2)=f_1^{TA(1)}(q^2)+\varepsilon
f_1^{TA(2)S}(q^2)+(1-\varepsilon)f_1^{TA(2)V}(q^2);
\end{equation}
\begin{eqnarray}
\label{eq:f1TA}
f_1^{TA(1)}(q^2)&=&-\int \frac{d^3p}{(2\pi)^3} \bar\Psi_F\left({\bf
    p}+\frac{2\epsilon_d}{E_F+M_F}{\bf
    \Delta}\right)\sqrt{\frac{\epsilon_Q(p)+m_Q}{2\epsilon_Q(p)}}\sqrt{\frac{\epsilon_q(p+\Delta)+m_q}{2\epsilon_q(p+\Delta)}}\cr
&&\times
\Biggl\{\frac{\epsilon_d}{E_F+M_F}\Biggl[\frac{M_F}{\epsilon_q(p+\Delta)+m_q}-\frac{M_I}{\epsilon_Q(p)+m_Q}-\frac{(M_I-M_F)\epsilon_d}{(\epsilon_q(p+\Delta)+m_q)(\epsilon_Q(p)+m_Q)}\Biggr]\cr
&&+ \frac{\bf p
  \Delta}{{\bf \Delta}^2}\Biggl[\frac{M_I+M_F}{\epsilon_Q(p)+m_Q}-\left(\frac{1}{\epsilon_q(p+\Delta)+m_q}
-\frac{1}{\epsilon_Q(p)+m_Q}\right)M_F\frac{E_F-M_F}{E_F+M_F}\cr
&&
-\frac{2\epsilon_d E_F}{(\epsilon_q(p+\Delta)+m_q)(\epsilon_Q(p)+m_Q)}\frac{E_F-M_F}{E_F+M_F}\Biggr]\cr
&&-\frac13\frac{M_I-M_F}{E_F+M_F}\frac{{\bf
    p}^2}{(\epsilon_q(p+\Delta)+m_q)(\epsilon_Q(p)+m_Q)}\Biggr\}\Psi_I({\bf
  p}); 
\end{eqnarray}

\begin{eqnarray}
  \label{eq:f1TAs}
f_1^{TA(2)S}(q^2)&=&\int \frac{d^3p}{(2\pi)^3} \bar\Psi_F\left({\bf
    p}+\frac{2\epsilon_d}{E_F+M_F}{\bf
    \Delta}\right)\sqrt{\frac{\epsilon_Q(p)+m_Q}{2\epsilon_Q(p)}}\sqrt{\frac{\epsilon_q(\Delta)+m_q}{2\epsilon_q(\Delta)}}\cr
&&\times\Biggl\{\frac{1}{2\epsilon_Q(\Delta)(\epsilon_Q(\Delta)+m_Q)}\frac{\bf p
  \Delta}{{\bf\Delta}^2}\left[M_I+M_F\left(1+\frac{E_F-M_F}{E_F+M_F}\right)\right]\cr
&&\times\left[M_I-\epsilon_Q(p)-\epsilon_d(p)\right]
-\frac{1}{2\epsilon_q(\Delta)(\epsilon_q(\Delta)+m_q)}\Biggl[\epsilon_q(\Delta)-m_q\cr
&&+(E_F-M_F)\left(1-\frac{\epsilon_d}{E_F+M_F}\right)-\frac{\bf p
  \Delta}{{\bf\Delta}^2}E_F\frac{E_F-M_F}{E_F+M_F}\Biggr]\cr
&&\times\Biggl[M_F-\epsilon_q\left({\bf
    p}+\frac{2\epsilon_d}{E_F+M_F}{\bf
    \Delta}\right) -\epsilon_d\left({\bf
   p}+\frac{2\epsilon_d}{E_F+M_F}{\bf\Delta}\right)\Biggr]\Biggr\}\Psi_I({\bf
   p});\ \ \ \ \
\end{eqnarray}

\begin{eqnarray}
  \label{eq:f1TAv}
f_1^{TA(2)V}(q^2)&=&-\int \frac{d^3p}{(2\pi)^3} \bar\Psi_F\left({\bf
    p}+\frac{2\epsilon_d}{E_F+M_F}{\bf
    \Delta}\right)\sqrt{\frac{\epsilon_Q(p)+m_Q}{2\epsilon_Q(p)}}\sqrt{\frac{\epsilon_q(\Delta)+m_q}{2\epsilon_q(\Delta)}}\cr
&&\times\Biggl\{\frac{1}{2\epsilon_Q(\Delta)(\epsilon_Q(\Delta)+m_Q)}\Biggl(\left[\epsilon_Q(\Delta)-m_Q+(E_F-M_F)\left(1-\frac{\epsilon_d}{E_F+M_F}\right)\right]\cr
&&\times\Biggl[\frac{\bf p
  \Delta}{{\bf\Delta}^2}\frac{1}{2E_d}\left(M_I+M_F\left(1+\frac{E_F-M_F}{E_F+M_F}\right)-\frac{\epsilon_d}{m_Q}\frac{E_F(M_I-E_F)}{E_F+M_F}\right)\cr
&&+\frac{\epsilon_d}{m_Q}\left(\frac{M_I}{E_F+M_F}+\frac{E_F-M_F}{E_F+M_F}\right)\Biggr]-\frac{\bf p
  \Delta}{{\bf\Delta}^2}\Biggl[M_I+M_F\left(1+\frac{E_F-M_F}{E_F+M_F}\right)\cr
&&-\frac{\epsilon_d}{m_Q}\frac{E_F(M_I-E_F)}{E_F+M_F}\Biggr]\Biggr)\left[M_I-\epsilon_Q(p)-\epsilon_d(p)\right]\cr
&&
+\frac{1}{2\epsilon_q(\Delta)(\epsilon_q(\Delta)+m_q)}\Biggl(\Biggl[\epsilon_q(\Delta)-m_q+(E_F-M_F)\left(1+\frac{\epsilon_d}{E_F+M_F}\right)\Biggr]\cr
&&\times\left[1-\frac{\epsilon_d}{m_q}\frac{M_F}{E_F+M_F}+\frac{\bf p
  \Delta}{{\bf\Delta}^2}\frac{E_F}{2E_d}\frac{E_F-M_F}{E_F+M_F}\left(1+\frac{\epsilon_d}{m_q}\frac{E_F+M_F}{E_F}\right)\right]\cr
&&-\frac{\bf p \Delta}{{\bf\Delta}^2}E_F\frac{E_F-M_F}{E_F+M_F}\left(1+\frac{\epsilon_d}{m_q}\frac{E_F+M_F}{E_F}\right)\Biggr)\cr
&&\times\Biggl[M_F-\epsilon_q\left({\bf
    p}+\frac{2\epsilon_d}{E_F+M_F}{\bf
    \Delta}\right) -\epsilon_d\left({\bf  p}+\frac{2\epsilon_d}{E_F+M_F}{\bf\Delta}\right)\Biggr]\Biggr\}\Psi_I({\bf p});
\end{eqnarray}

\begin{equation}
f_2^{TA}(q^2)=f_2^{TA(1)}(q^2)+\varepsilon
f_2^{TA(2)S}(q^2)+(1-\varepsilon)f_2^{TA(2)V}(q^2);
\end{equation}
\begin{eqnarray}
\label{eq:fTA2}
f_2^{TA(1)}(q^2)&=&-\int \frac{d^3p}{(2\pi)^3} \bar\Psi_F\left({\bf
    p}+\frac{2\epsilon_d}{E_F+M_F}{\bf
    \Delta}\right)\sqrt{\frac{\epsilon_Q(p)+m_Q}{2\epsilon_Q(p)}}\sqrt{\frac{\epsilon_q(p+\Delta)+m_q}{2\epsilon_q(p+\Delta)}}\cr
&&\times\Biggl\{1+\frac{\epsilon_d}{E_F+M_F}\Biggl[\frac{M_I-M_F}{M_I}\left(\frac{M_F}{\epsilon_q(p+\Delta)+m_q}-\frac{M_I}{\epsilon_Q(p)+m_Q}\right)\cr
&&
+(E_F-M_F)\left(\frac1{\epsilon_q(p+\Delta)+m_q}+\frac{1}{\epsilon_Q(p)+m_Q}\right)\cr
&&
-\frac{\epsilon_d(E_F+M_F)}{(\epsilon_q(p+\Delta)+m_q)(\epsilon_Q(p)+m_Q)}\left(
   \frac{(M_I-M_F)^2}{M_I(E_F+M_F)}-
   \frac{E_F-M_F}{E_F+M_F}\right)\Biggr]\cr
&&+\frac{\bf
  p \Delta}{{\bf
   \Delta}^2}\Biggl[\frac{M_I-M_F}{M_I}\left(\frac{M_F}{\epsilon_q(p+\Delta)+m_q}+\frac{M_I}{\epsilon_Q(p)+m_Q}\right)\cr
&&-\left(\frac1{\epsilon_q(p+\Delta)+m_q}-\frac{1}{\epsilon_Q(p)+m_Q}\right)(E_F+M_F)\cr
&&\times\left(\frac{M_I-M_F}{2M_I}\left(1-\frac{E_F-M_F}{E_F+M_F}\right)\left(1+\frac{E_F-M_F}{E_F+M_F}\right)+\left(\frac{E_F-M_F}{E_F+M_F}\right)^2\right)\cr
&&-\frac{2\epsilon_d(M_I-M_F)E_F}{M_I(\epsilon_q(p+\Delta)+m_q)(\epsilon_Q(p)+m_Q)}\frac{E_F-M_F}{E_F+M_F}\Biggr]\cr
&& +\frac{{\bf
    p}^2}{(\epsilon_q(p+\Delta)+m_q)(\epsilon_Q(p)+m_Q)}\left[1+\frac23\frac{E_F-M_F}{E_F+M_F}-\frac13\frac{(M_I-M_F)^2}{M_I(E_F+M_F)}\right]
\Biggr\}\Psi_I({\bf p});\cr&&
\end{eqnarray}

\begin{eqnarray}
  \label{eq:f2TAs}
f_2^{TA(2)S}(q^2)&=&-\int \frac{d^3p}{(2\pi)^3} \bar\Psi_F\left({\bf
    p}+\frac{2\epsilon_d}{E_F+M_F}{\bf
    \Delta}\right)\sqrt{\frac{\epsilon_Q(p)+m_Q}{2\epsilon_Q(p)}}\sqrt{\frac{\epsilon_q(\Delta)+m_q}{2\epsilon_q(\Delta)}}\cr
&&\times\Biggl\{\frac{1}{2\epsilon_Q(\Delta)(\epsilon_Q(\Delta)+m_Q)}\Biggl(\left[\epsilon_Q(\Delta)-m_Q+(E_F-M_F)\left(1-\frac{\epsilon_d}{E_F+M_F}\right)\right]\cr
&&-\frac{\bf p
  \Delta}{{\bf\Delta}^2}\left[\frac{M_I^2-M_F^2}{M_I}+\left(E_F-\frac{M_F^2}{M_I}\right)\frac{E_F-M_F}{E_F+M_F}\right]\Biggr)\left[M_I-\epsilon_Q(p)-\epsilon_d(p)\right]\cr
&&
-\frac{1}{2\epsilon_q(\Delta)(\epsilon_q(\Delta)+m_q)}\Biggl[\frac{M_F}{M_I}\left(\epsilon_q(\Delta)-m_q+(E_F-M_F)\left(1-\frac{\epsilon_d}{E_F+M_F}\right)\right)\cr
&&+\frac{\bf p
  \Delta}{{\bf\Delta}^2}(E_F-M_F)\left(1+\frac{E_F}{M_I}\frac{M_I+M_F}{E_F+M_F}\right)\Biggr]\cr
&&\times\Biggl[M_F-\epsilon_q\left({\bf
    p}+\frac{2\epsilon_d}{E_F+M_F}{\bf
    \Delta}\right) -\epsilon_d\left({\bf  p}+\frac{2\epsilon_d}{E_F+M_F}{\bf\Delta}\right)\Biggl]\Psi_I({\bf p});
\end{eqnarray}

\begin{eqnarray}
  \label{eq:f2TAv}
f_2^{TA(2)V}(q^2)&=&-\int \frac{d^3p}{(2\pi)^3} \bar\Psi_F\left({\bf
    p}+\frac{2\epsilon_d}{E_F+M_F}{\bf
    \Delta}\right)\sqrt{\frac{\epsilon_Q(p)+m_Q}{2\epsilon_Q(p)}}\sqrt{\frac{\epsilon_q(\Delta)+m_q}{2\epsilon_q(\Delta)}}\cr
&&\times\Biggl\{\frac{1}{2\epsilon_Q(\Delta)(\epsilon_Q(\Delta)+m_Q)}\Biggl(\left[\epsilon_Q(\Delta)-m_Q+(E_F-M_F)\left(1-\frac{\epsilon_d}{E_F+M_F}\right)\right]\cr
&&\times\Biggl[1+\frac{\epsilon_d}{m_Q}\left(\frac{M_I-M_F}{E_F+M_F}-\frac{M_F}{M_I}\frac{E_F-M_F}{E_F+M_F}\right)+\frac{\bf p \Delta}{{\bf\Delta}^2}\frac{1}{2E_d}\Biggl(\frac{M_I^2-M_F^2}{M_I}\cr
&&+\left(E_F-\frac{M_F^2}{M_I}\right)\frac{E_F-M_F}{E_F+M_F}-\frac{\epsilon_d}{m_Q}\frac{E_F}{E_F+M_F}\left(\frac{M_I^2+M_F^2}{M_I}-E_F+M_F\right)\Biggr) \Biggr]
\cr
&&-\frac{\bf p
  \Delta}{{\bf\Delta}^2}\Biggl[\frac{M_I^2-M_F^2}{M_I}+\left(E_F-\frac{M_F^2}{M_I}\right)\frac{E_F-M_F}{E_F+M_F}\cr
&&-\frac{\epsilon_d}{m_Q}\frac{E_F}{E_F+M_F}\left(\frac{M_I^2+M_F^2}{M_I}-E_F+M_F\right)\Biggr]\Biggr)\left[M_I-\epsilon_Q(p)-\epsilon_d(p)\right]\cr
&&
-\frac{1}{2\epsilon_q(\Delta)(\epsilon_q(\Delta)+m_q)}\Biggl[\left(\epsilon_q(\Delta)-m_q+(E_F-M_F)\left(1-\frac{\epsilon_d}{E_F+M_F}\right)\right)
\cr
&&\times\Biggl(\frac{M_F}{M_I}\left[1+\frac{\epsilon_d}{m_q}\frac{M_F}{E_F+M_F}\left(\frac{E_FM_I}{M_F^2}-1\right)\right]+\frac{\bf
   p \Delta}{{\bf\Delta}^2}\frac{E_F-M_F}{2E_d}
\cr
&&\times\left(1+\frac{E_F}{M_I}\frac{M_I+M_F}{E_F+M_F}-\frac{\epsilon_d}{m_q}\left[1-\frac{M_F}{M_I}\left(1+\frac{E_F-M_F}{E_F+M_F}\right)\right]\right) \Biggr)
\cr
&&+\frac{\bf p
  \Delta}{{\bf\Delta}^2}(E_F-M_F)\left(1+\frac{E_F}{M_I}\frac{M_I+M_F}{E_F+M_F}-\frac{\epsilon_d}{m_q}\left[1-\frac{M_F}{M_I}\left(1+\frac{E_F-M_F}{E_F+M_F}\right)\right]\right)\Biggr]\cr
&&\times\Biggl[M_F-\epsilon_q\left({\bf
    p}+\frac{2\epsilon_d}{E_F+M_F}{\bf
    \Delta}\right) -\epsilon_d\left({\bf  p}+\frac{2\epsilon_d}{E_F+M_F}{\bf\Delta}\right)\Biggl]\Psi_I({\bf p});
\end{eqnarray}

where 
\[ \left|{\bf \Delta}\right|=\sqrt{\frac{(M_{I}^2+M_F^2-q^2)^2}
{4M_{I}^2}-M_F^2},\]
superscripts (1) and (2) correspond to vertex functions $\Gamma^{(1)}$
and $\Gamma^{(2)}$,  $S$ and $V$ correspond to the scalar and vector confining
potentials, $\epsilon_d$ is the diquark energy,
\[ E_F=\sqrt{M_F^2+{\bf \Delta}^2},\quad \epsilon_{q,d}(
\Delta)=\sqrt{m_{q,d}^2+{\bf\Delta}^2}, \quad
 \epsilon_{q,d}(p+\lambda
\Delta)=\sqrt{m_{q,d}^2+({\bf p}+\lambda{\bf \Delta})^2} \quad (q=b,s), \]
subscripts $I$ and $F$ denote the initial $\Lambda_b$ and final $\Lambda$ baryons.

\end{document}